\documentclass{emulateapj}

\begin{document}

\title{Anisotropic AGN Outflows and Enrichment of the Intergalactic Medium}

\author{Jo\"el Germain\altaffilmark{1,2}, 
Paramita Barai\altaffilmark{1,2}, and 
Hugo Martel\altaffilmark{1,2}}

\altaffiltext{1}{D\'epartement de physique, de g\'enie physique et d'optique,
Universit\'e Laval, Qu\'ebec, QC, Canada}

\altaffiltext{2}{Centre de Recherche en Astrophysique du Qu\'ebec}

\begin{abstract}

We investigate the cosmological-scale influence of outflows driven
by AGNs
on metal enrichment of the intergalactic medium.
AGNs are located in dense cosmological structures which
tend to be anisotropic.
We designed a semi-analytical model for anisotropic AGN outflows
which expand away along the direction of least resistance. This model was
implemented into a cosmological numerical simulation algorithm
for simulating the growth of large-scale structure in the universe.
Using this modified algorithm, we perform a series of
9 simulations inside cosmological volumes of size $(128~h^{-1}{\rm Mpc})^3$,
in a concordance $\Lambda$CDM universe, varying the opening angle of the
outflows, the lifetimes of the AGNs, their kinetic fractions, and
their level of clustering.
For each simulation, we compute the volume fraction of the IGM enriched 
in metals by the outflows. 
The resulting enriched volume fractions are relatively small at 
$z \gtrsim 2.5$, and then grow rapidly afterward up to $z = 0$. 
We find that AGN outflows enrich from 65\% to 100\% of the 
entire universe at the present epoch, 
for different values of the model parameters. 
The enriched volume fraction depends weakly
on the opening angle of the outflows. However,
increasingly anisotropic outflows preferentially enrich underdense regions, 
a trend found more prominent at higher redshifts and decreasing at lower 
redshifts. The enriched volume fraction increases with increasing
kinetic fraction and decreasing AGN lifetime and level of clustering.

\end{abstract}


\section{INTRODUCTION} 
\label{sec-intro} 

Active galaxies are believed to be powered by accretion of matter onto the 
central supermassive black holes (SMBHs) \citep[e.g.,][]
{kormendy95, ferrarese05}, 
liberating enormous amounts of energy 
(often in the forms of ejected energetic jets/outflows), 
affecting their environments from pc to Mpc scales. 
Active Galactic Nuclei (AGN) are believed to influence the formation 
and evolution of galaxies and large-scale structures over the cosmic epoch, 
in the form of feedback, whereby 
the overall properties of a galaxy can be regulated by its central BH 
\citep[e.g.,][]{silk98, king03, wyithe03, granato04, murray05, begelman05, 
pipino09, ciotti09}. 
Recent concordance models of galaxy formation through hierarchical clustering 
in the cold dark matter cosmology invoke feedback from AGN 
to explain several observations \citep[for a review]{best07}, such as 
the central SMBH - host galaxy bulge correlations 
\citep[e.g.,][]{magorrian98, gebhardt00, mcLure02} and
the sharp cutoff at the bright end of the galaxy luminosity function. 

During the quasar era ($1<z<3$), 
the star-formation rate, the comoving density of AGN, and the 
merger rate of galaxies, 
are all observed to reach their peak values. 
Such a common trend points to a coevolution scenario: galaxies and AGN 
are argued to have evolved together influencing each other's growth. 
\citet{polletta08} discovered two sources at $z\sim3.5$ exhibiting 
both powerful starburst and AGN activities, 
which is interpreted as a coevolution phase of star formation and AGN 
as predicted by various models of galaxy formation and evolution. 
Observations of an AGN-starburst galaxy by \citet{croston08} show that AGN 
feedback has a substantial impact on the galaxy and its surroundings. 
Cosmological simulations by \citet{diMatteo05} indicate that 
the energy released from an accreting SMBH 
can halt further accretion onto the BH and drive away gas, 
self-regulating galaxy growth by shutting off the quasar and quenching 
star-formation in the galaxy, and thus reddening it rapidly. 

The nature of feedback from AGN on star/galaxy formation can be 
either negative (quenching) or positive (enhancement),  
as shown by different studies. The exact role played by AGNs
possibly depends on one or more factors, such as radio-loudness, 
jet power, lifetime/duty cycle, environmental factors like ambient 
gas density, etc. 
In galaxy clusters, AGN outflows are believed to stop the cooling flow 
and heat up the intracluster medium 
by giant cavities, buoyant bubbles and shock fronts 
\citep[e.g.,][]{mcNamara07}. 
At the same time, substantial star-formation rates are observed in the 
central radio galaxies 
of cooling flow clusters triggered by the radio source \citep{mcNamara02}. 

Some observations \citep[e.g.,][]{schawinski07} indicate that AGN quench 
star-formation in 
their host galaxies at $z < 2$ turning them red and dead, 
in the process generating the observed bimodal color distribution of galaxies. 
Hydrodynamical simulations of major mergers of equal- and unequal-mass 
disk and elliptical galaxies show that BH feedback can 
terminate star-formation \citep{springel05, johansson08}. 
\citet{antonuccio-Delogu08} 
analyzed the different physical factors impacting the star-formation rate, and 
found that there is suppression of star-formation in galaxies from 
AGN jet-induced feedback. 
Semi-analytic models \citep{bower06, croton06}
show that feedback due to AGN can quench cooling flows and 
star formation in massive halos,
explaining the observed downsizing in the
stellar populations of galaxies \citep[e.g.,][]{juneau05}.  

Shock-induced and jet-induced star formation in radio galaxies can
explain the radio-optical alignment observations. 
In hydrodynamic simulations including radiative cooling, 
AGN-driven shock waves compress a gas cloud, 
causing it to break up into multiple smaller, dense fragments, 
which survive for many dynamical timescales, turn Jeans unstable, 
and form stars \citep[e.g.,][]{mellema02, fragile04, vanBreugel04}. 
Expanding radio jets/lobes generate shocks which propagate through 
inhomogeneous clumpy ambient gas clouds and
trigger gravitational collapse of overdense ambient gas clouds, 
and star formation 
\citep[e.g.,][]{begelman89, deYoung89, rees89, bicknell00}. 

Studies claim to have found direct observational evidence for AGN 
feedback at $z\sim2$. 
\citet{nesvadba08a}, using VLT data, identified kpc-sized 
outflows of ionized gas in $z \sim 2 - 3$ radio galaxies. 
These bipolar outflows have the expected signatures of being powerful 
AGN-driven winds, 
energetic enough to terminate star formation in the massive host galaxies. 
At the same time there have been few observations 
\citep[e.g.,][]{krongold07, schlesinger09} 
implying that quasar winds are unlikely to produce 
important evolutionary effects on their larger environments, 
which poses a challenge for the scenario of strong AGN feedback in 
galaxy evolution. 

A fraction of AGN are radio loud, and there are claims that the radio-loudness 
depends on SMBH mass and Eddington ratio \citep[e.g.,][]{laor00, rafter09}. 
The lifetimes of radio sources ($\sim 10^6 - 10^8$ yrs) are observationally 
inferred to be significantly shorter than the ages of their host galaxies,
suggesting that the hosts cycle between radio-loud and radio-quiet phases. 
AGN feedback is a natural explanation: 
the hot ISM in a galaxy episodically cools to fuel the central AGN, 
which triggers radio activity, and then the emanated radio jets reheat the gas 
\citep{ciotti07}. 

A large fraction of AGN are observed to host outflows, 
in a wide variety of forms \citep[see][for reviews]{crenshaw03, begelman04, 
everett07}. 
Radio galaxies with collimated relativistic jets and/or huge overpressured 
cocoons 
emitting radio synchrotron radiation constitute $\sim 10\%$ of all quasars 
\citep{peterson97}. 
Blue-shifted broad absorption lines (BALs) in the UV and optical are seen in 
an additional $\sim 10 - 15\%$ of QSOs \citep[e.g.,][]{reichard03}. 
Many BAL and other quasars in SDSS exhibit highly-ionized broad blue-shifted 
emission lines of e.g., C IV \citep[e.g.,][]{richards02}. 
Intrinsic absorption lines in the UV have been detected in $> 50 \%$ 
of Seyfert I galaxies studied by \citet{crenshaw99}. 
Outflows have been observed in [OIII] emission lines 
in nearby Seyfert galaxies \citep[e.g.,][]{das05}. 
Warm absorbers seen in X-rays indicate ionized outflows in Seyferts and QSOs 
\citep[e.g.,][]{blustin05, krongold07}. 
More high-velocity highly-ionized outflows have been detected as 
absorption lines 
in X-rays \citep[e.g.,][]{chartas03, pounds03, dasgupta05, oBrien05}.
The full mechanism of AGN feedback and how it operates on different scales 
is poorly understood, both in theoretical and observational aspects. 
In this paper we investigate the impact of energetic outflows emanating from 
AGN on large-scale cosmological volumes. 

There have been previous studies on the cosmological impact 
of quasar outflows at large scales 
(\citealt{FL01}, hereafter FL01; \citealt{so04}, hereafter SO04; 
\citealt{lg05}, hereafter LG05). 
\citet{barai08} used cosmological simulations to
investigate the large-scale influence of radio galaxies over the Hubble time. 
All these studies have considered outflows expanding with a spherical geometry.
However, in realistic cosmological scenarios, where the density distribution 
show significant structures in the form of filaments, pancakes, etc., 
outflows are expected to expand anisotropically on large 
scales. 
Recent observations support the picture of anisotropic bipolar 
outflows on different scales. 
\citet{nesvadba08} found spectroscopic evidence for bipolar outflows in 
three powerful radio galaxies at $z \sim 2 - 3$, with kinetic energies 
equivalent to $0.2\%$ of the rest-mass of the SMBH. 
These outflows possibly indicate a significant phase in the
evolution of the host galaxy. 
On small scales, \citet{anathpindika08} showed that the observed 
outflows from young stellar objects tend to be orthogonal to the filaments 
that contain their driving sources. 
Numerical simulations on galaxy-scales \citep[e.g.,][]{
mf99,recchi09} show that 
bipolar winds and outflows
expand preferentially along the steepest density slope, 
i.e., the direction perpendicular to the galaxy plane, while simulations
at larger scales \citep{martel01a,martel01b} show that outflows emerging
from cosmological structures expand preferentially along the
direction of least resistance. 

The goal of our work is to study the global large-scale influence of 
outflows from AGN 
on the intergalactic medium (IGM) in a cosmological context.
We have designed a semi-analytical model for anisotropic outflows,
which we implemented into a numerical algorithm for cosmological
simulations. Using this algorithm, we simulate the propagation of
AGN-driven outflows into the IGM, in a $\Lambda$CDM universe. 
We then explore the large-scale impact 
of the cosmological population of AGN outflows over the age of
the universe.
In this paper, we focus on metal enrichment of the IGM, 
estimate the volume fraction of the universe enriched by AGN outflows
as a function of redshift, and its dependence on the various parameters
of the model. In a forthcoming paper, we will
will focus on the metal content and metallicity of the IGM and its
observational consequences.

This paper is organized as follows. In
\S\ref{sec-numerical} we describe our semi-analytical model for
anisotropic outflows, and its implementation into a cosmological
N-body simulation algorithm.
The results are presented and discussed in \S\ref{sec-results}. 
We present our conclusions in \S\ref{sec-conclusion}. 

\section{THE NUMERICAL METHOD} 
\label{sec-numerical} 

Our numerical method consists of three ingredients, (1) a cosmological
Particle-Mesh algorithm for simulating the formation and evolution
of large-scale structures in the universe, (2) a model for the masses,
formation epochs, lifetimes, and spatial distributions
of AGNs, and (3) a semi-analytical model for the propagation
of anisotropic outflows and the resulting metal-enrichment
of the IGM. We discuss these various ingredients in the following 
subsections.

\subsection{The PM Algorithm}
\label{sec-num-PM} 

We simulate the growth of large-scale structure in a cubic cosmological
volume of comoving size $L_{\rm box} = 128~h^{-1}$ Mpc $= 182.6\,{\rm Mpc}$
with periodic boundary conditions, 
using a Particle-Mesh (PM) algorithm \citep{he88}. We use $256^3$ equal mass
particles and a $512^3$ grid. This corresponds to a particle mass
$m_{\rm part}=1.38\times10^{10} M_\odot$, 
and a grid spacing $\Delta=0.357\,{\rm Mpc}$. 
Note that this length resolution is sufficient for our purpose; 
we do not need the extra length resolution that a $\rm P^3M$ algorithm 
would provide, and using PM instead of $\rm P^3M$ results in a 
major speed-up of the calculation. LG05 also used a PM algorithm
for their study of AGN outflows. 

We consider a $\Lambda$CDM model with a present 
baryon density parameter $\Omega_{b,0}=0.0462$, 
total matter (baryons + dark matter) density parameter $\Omega_0=0.279$, 
cosmological constant $\Omega_{\Lambda,0}=0.721$, 
Hubble constant $H_0=70.1\rm\,km\,s^{-1}Mpc^{-1}$ ($h=0.701$),
primordial tilt $n_s=0.960$, and CMB temperature $T_{\rm CMB}=2.725$, 
consistent with the results of {\sl WMAP5} 
combined with the data from baryonic acoustic oscillations and supernova
studies 
\citep{hinshaw08}. We generate initial conditions at
redshift $z=24$, and evolve the cosmological volume up 
to a final redshift $z=0$.

\subsection{Distribution in Redshift and Luminosity} 
\label{sec-num-QLF} 

\begin{figure} 
\includegraphics[width = \linewidth]{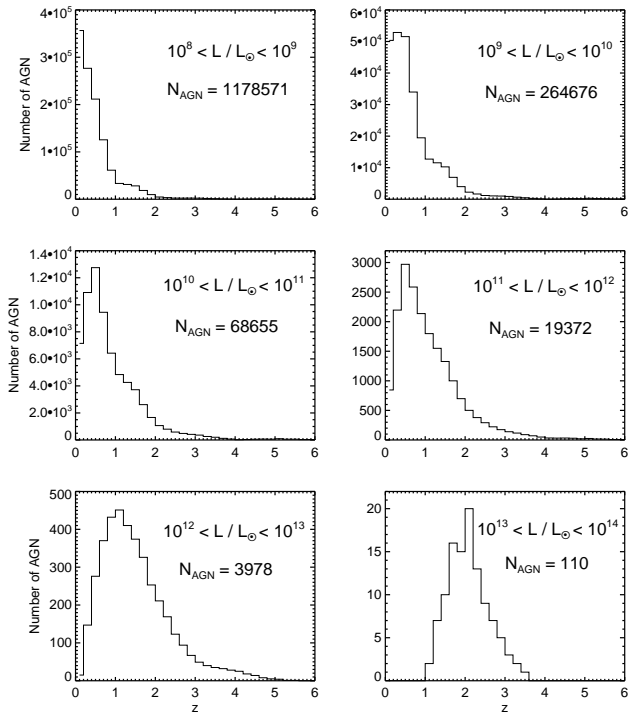}
\caption{ 
Redshift distribution of a total of $1\,535\,362$ 
sources generated according to 
the QLF (\S\ref{sec-num-QLF}) in various luminosity bins, 
with parameter values: $T_{\rm AGN} = 100$ Myr, and $f_{\rm outflow}= 0.6$.
The number of sources $N_{\rm AGN}$ in each luminosity bin is
indicated.}
\label{fig1} 
\end{figure} 

We determine the luminosities and birth times of the cosmological AGN 
population 
by adopting the redshift-dependent luminosity distribution of AGNs from 
the work 
of \citet{hopkins07} on bolometric quasar luminosity function (QLF). 
The QLF is expressed as a standard double power law,
\begin{equation}
\phi(L,z) \equiv \frac{d\Phi}{d \log L} 
= \frac{\phi_{\star}} {(L/L_{\star})^{\gamma_1} + (L/L_{\star})^{\gamma_2}}, 
\end{equation}

\noindent 
which gives the number of quasars per unit comoving volume, 
per unit log luminosity interval. 
The values of the amplitude $\phi_{\star}$, break luminosity $L_{\star}$, 
and the slopes, $\gamma_1$ and $\gamma_2$ evolve with redshift. We use
the values given in Table 2 of \citet{hopkins07}, which cover the redshift 
range $z=6.0-0.1$. For a given redshift $z$, we get the luminosity 
distribution by interpolating between two consecutive lines
in that table. For redshifts $z<0.1$, we simply use the luminosity function
at $z=0.1$. 

We assume that a fraction $f_{\rm outflow}= 0.6$ of AGNs produce outflows 
\citep{ganguly08}.
The number of AGNs with outflows in the simulation box of comoving volume 
$V_{\rm box} = L_{\rm box}^3$, 
and within luminosity interval $[L, L+dL]$ at redshift $z$ is given by 
\begin{equation}
N(L,z) = f_{\rm outflow}V_{\rm box}\phi(L,z)d[\log L]\,.
\label{nlz}
\end{equation}

\noindent 
We assume that AGNs have luminosities between
$L_{\min} = 10^{8} L_{\odot}$ 
and $L_{\max} = 10^{14} L_{\odot}$ \citep{crenshaw03}. 
The fiducial value for the AGN activity lifetime is taken as 
$T_{\rm AGN} =10^8$ yr, with other values considered in \S\ref{sec-Tagn}. 

We calculate the number and birth time of the AGNs as follows. 
We divide the luminosity range $L=[L_{\min},L_{\max}]$ in bins
of size $\Delta\log(L/L_{\odot})=0.1$.
For each luminosity bin,
we start at redshift $z=z_{\max}=6$, and calculate the number of
AGNs with outflows in that bin at that redshift using equation~(\ref{nlz}).
We then assign to each AGN a random age $t_{\rm age}$ between 0 
(AGN just being born)
and $T_{\rm AGN}$ (AGN just about to die), and calculate the
birth redshift $z_{\rm birth}=z(t_{\max}-t_{\rm age})$, where $t_{\max}$
is the cosmic time corresponding to redshift $z_{\max}$. We then move
forward in time by intervals of $\Delta t=10\,\rm Myr$.\footnote{The 
length of this interval does not matter as 
long as it is sufficiently smaller than $T_{\rm AGN}$.}
For each new time~$t$,
we calculate the number of AGNs with outflows
using equation~(\ref{nlz}), subtract from
that number the number of AGNs present at earlier times that are still 
alive at that time, and assign to each new AGN an age between 0 and
$\Delta t$ (since, if that AGN was not present at the previous time, it must
have appeared during the last interval $\Delta t$).

Using the QLF, we obtain the entire cosmological population of AGNs
in the simulation box starting from $z_{\max}$, namely the birth 
redshift ($z_{\rm bir}$), 
switch-off redshift ($z_{\rm off}$) and bolometric luminosity 
($L_{\rm bol}$) of each source. 
Figure~\ref{fig1} shows the redshift distribution of the total population of 
$1\,535\,362$ sources produced using $T_{\rm AGN} =10^8$ yr. 

\subsection{Spatial Location of AGNs}
\label{sec-num-locate} 

By far the AGNs have been observed \citep[e.g.,][]{dodorico08} 
to be hosted in high density regions of the universe. 
In order to determine the spatial location of the AGNs 
in our simulations, we consider the density distribution
on the $512\times512\times512$ grid, as calculated by the PM code.
We filter this density using a gaussian filter containing 
a mass 
$10^{10} M_{\odot}$ \citep[e.g.,][]{kauffmann03, hickox09}, 
considering that as the minimum mass of a galaxy that might host an AGN 
(for details of the filtering technique, we refer the reader to \S5 of 
\citealt{martel05}). 
We then identify all grid points where the value of the filtered density 
exceeds the values at the 26 neighboring grid points. 
These are the locations of the density peaks. 

At each timestep of the simulation, we spatially locate the new AGNs 
born during that epoch (\S\ref{sec-num-QLF}, 
whose $z_{\rm bir}$ values fall within the timestep interval) 
at the local density peaks in the cosmological volume. 
We should have enough peaks to assign to each new AGN 
an unique location, but 
at the same time we should locate the AGNs preferentially
in high-density regions. 
So we adopt a redshift-dependent limiting density $\rho_{\rm lim}^{\phantom1}$
to select potential peaks for locating the sources. 
We consider the peaks that have a filtered density 
$\rho>\rho_{\rm lim}^{\phantom1}$, 
and each new AGN is located at the center of one such peak, 
selected randomly but excluding peaks already containing an AGN. 
The number of AGN increases with time, hence we need to
reduce the value of $\rho_{\rm lim}$ with redshift in order to select
a sufficient number of peaks.
We use 
$\rho_{\rm lim}^{\phantom1}=10\bar{\rho}$ when $z > 3$, 
$\rho_{\rm lim}^{\phantom1}=7\bar{\rho}$ when $3 \ge z > 2$, 
$\rho_{\rm lim}^{\phantom1}=5\bar{\rho}$ when $2 \ge z > 1$, 
$\rho_{\rm lim}^{\phantom1}=3\bar{\rho}$ when $1 \ge z > 0.75$, and 
$\rho_{\rm lim}^{\phantom1}=2\bar{\rho}$ when $z \le 0.75$, 
where $\bar{\rho}(z)=3\Omega_0H_0^2(1+z)^3/8\pi G$ is the 
mean density of the universe at that redshift.

Note that our method for locating AGNs differs from the one
used by LG05. In their simulations, AGNs could be located
anywhere in the computational volume, with a probability that
depended on the local density. We have to restrict the potential
locations of AGNs to local density peaks, to be consistent with our
anisotropic outflow model (see below). This is not a severe restriction,
because we do expect massive galaxies that host AGNs to form 
at the locations of density peaks anyway. 
We consider alternate methods of locating the AGNs in \S\ref{sec-bias}. 

\subsection{Anisotropic Outflows} 
\label{sec-num-aniso} 

\citet{pmg07} (hereafter PMG07)
have developed an anisotropic outflow model, in which outflows
expanding into an anisotropic medium follows the path of least resistance.
The model was applied to supernovae-driven outflows generated
by low-mass galaxies. In this paper we apply the same model to AGN-driven
outflows generated by massive galaxies. The model is 
essentially the same, except for some additional terms 
in the equations driving the outflow. 
We refer the reader to PMG07 for details. 

\begin{figure}
\centering 
\includegraphics[width = \linewidth]{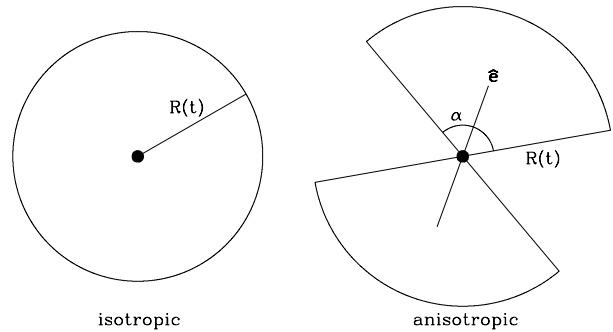}
\caption{Geometry of the outflow. $R(t)$ and $\alpha$ are the radius
and opening angle of the outflow, respectively. The unit vector ${\bf\hat e}$ 
indicates the direction of least resistance. Figure taken from PMG07.}
\label{geometry}
\end{figure}

The outflow is approximated as two expanding bipolar cones with opening
angle $\alpha$, expanding radially in opposite directions, as shown in
Figure~\ref{geometry}. The
opening angle is treated as a free parameter. The limits
$\alpha=\pi$ and $\alpha\ll1$ correspond to the cases of isotropic outflows 
and jets, respectively. Note that in a given simulation, we use the same value 
of $\alpha$ for all outflows. The total volume of the outflow is 
$4 \pi R^3(1-\cos\alpha/2)/3$.

We assume that the outflow will follow the direction of least resistance
as it travels away from the source. Our technique for determining this
direction was presented in PMG07 and \citet{grenon07}. 
Essentially, we perform a second-order Taylor expansion of the density
around each peak where a source is located:
\begin{equation}
\rho(x,y,z)=\rho_{\rm peak}-Ax^2-By^2-Cz^2-2Dxy-2Exz-2Fyz\,,
\label{taylor1}
\end{equation}

\noindent where the coordinates $x$, $y$, $z$ are measured from the position
of the peak. Since the peak is by definition a local maximum, there are
no linear terms in the expansion. In practice, we determine the coefficients
$A$ through $F$ by performing a least-square fit of equation~(\ref{taylor1})
to all the grid points within a distance 
$2 \Delta$ from the peak ($\Delta$ being the grid spacing). 
We then rotate the coordinate axes such that, 
in the new coordinate system $(x',y',z')$, the cross-terms vanish:
\begin{equation}
\rho(x',y',z')=\rho_{\rm peak}-A'{x'}^2-B'{y'}^2-C'{z'}^2\,.
\label{taylor2}
\end{equation}

\noindent The three coefficients $A'$, $B'$, $C'$ are always positive,
otherwise we would not have a peak, but rather a local minimum
or a saddle point. The largest of these coefficients gives us the direction
along which the density drops the fastest as we
move away from the peak. We take it as the direction of least resistance.

To obtain a complete description of the outflow, we still need to provide
an outflow model that will give us the time-dependent radius $R(t)$. 
This is where our model differs from the one of PMG07, which is designed
for supernovae-driven outflows. We describe
our outflow model in the next subsection.

\subsection{Outflow Model}
\label{sec-num-outflow} 

Despite the observational differences between various kinds of outflows, 
the important point relevant for the present study is that the AGNs 
hosting outflows 
form a random subset of the whole AGN population (SO04). 
For simplicity we assume the same outflow model 
for all kinds of AGN, as in FL01, SO04, and LG05. 
Neglecting outflow formation timescales (since we are bound by 
simulation resolutions), 
we consider that each AGN produces an outflow right from its birth. 
The outflow expands into the IGM 
with an anisotropic geometry, getting channeled into low-density regions. 

We set the kinetic luminosity $L_K $ carried by the AGN jets equal to
a constant fraction of the bolometric luminosity, 
\begin{equation}
\label{lagn}
L_K = \epsilon_K L_{\rm bol}, 
\end{equation} 

\noindent
where $\epsilon_K$ is the kinetic fraction.
For our initial simulations, we use the value $\epsilon_K = 0.1$ 
(\citealt{willott99}; FL01; LG05; \citealt{chartas07, shankar08}), 
which is probably an upper limit for BAL outflows \citep{nath02},
but we consider other values of $\epsilon_K$ in \S\ref{sec-epsK}. 
SO04 used $\epsilon_K = 0.05$. 
The total kinetic energy transported by the jets
during an AGN's active lifetime, $E_K = L_K T_{\rm AGN}$,
is assumed to be converted to the thermal energy of the outflow. 

We assume that the central black hole (of mass $M_{\rm BH}$) radiates at 
the Eddington limit, and provides the AGN bolometric luminosity, 
\begin{eqnarray}
L_{\rm bol} & = & 
L_{\rm Eddington} = \frac{4\pi Gcm_p}{\sigma_e}M_{\rm BH} \nonumber \\ 
& = & 1.26\times10^{38}{\rm erg\,s^{-1}}\left(\frac{M_{\rm BH}}{M_{\odot}}\right). 
\end{eqnarray}

\noindent
Then using the ratio of central BH mass to the galaxy bulge stellar mass, 
$M_{\rm BH} / M_{\rm bulge} = 0.002$ 
(e.g., \citealt{magorrian98}; \citealt{gebhardt00}; FL01; 
\citealt{marconi03}), 
and scaling it with the universal ratio of the density of matter to baryons, 
we obtain the total (baryonic+dark matter) mass of the galaxy hosting the AGN, 
$M_{\rm gal} = (\Omega_0/\Omega_{b,0}) M_{\rm bulge}$. 

The model of a spherical bubble expanding as a thin shell (of mass $M_s$) 
in a cosmological volume \citep[e.g.,][]{tegmark93} is used to obtain 
the radius, $R$, of the outflow. 
The rate at which IGM mass is swept up by an anisotropic outflow is given by
\begin{equation} 
\dot{M_s}=\cases{
0\,, & $v_p\geq\dot{R}\,;$ \cr
4\pi R^2\left[1-\cos\left(\alpha/2\right)\right]
\rho_x\left[\dot{R}-v_p\right]\,, & $v_p<\dot{R}\,;$ \cr}
\end{equation} 

\noindent
where $\rho_x$ is the density of the external gas, and 
$v_p$ is the gas velocity due to infall onto the dark matter halo 
hosting the AGN. As a simplifying approximation, we consider that the gas
density is equal to
the mean baryon density of the universe at the corresponding redshift, 
\begin{equation}
\label{rhox} 
\rho_x(z)=\frac{3 H^2(z)\Omega_b(z)}{8\pi G} 
=\frac{3 H_0^2\Omega_{b,0}}{8\pi G}(1+z)^3. 
\end{equation} 

\noindent
For the gas infall velocity, we assume $v_p = H(z) R$, 
where $H(z)$ is the Hubble constant at redshift $z$. 
This is simply the Hubble expansion of a cosmological volume. 

After correcting for anisotropic expansion, the acceleration of the 
shell can be written as, 
\begin{eqnarray}
\label{eq-late-R}
\ddot{R} & = & 
\frac{4\pi R^2\left[1-\cos\left(\alpha/2\right)\right]}{M_s}
(p^{\phantom1}_T+p^{\phantom1}_B-p^{\phantom1}_x) \nonumber \\ 
& & -\frac{G}{R^2}\left[M_d(R)+M_{\rm gal}+\frac{M_s}{2}\right] \nonumber \\ 
& & +\Omega_\Lambda(z)H^2(z)R-\frac{\dot{M_s}}{M_s}\left[\dot{R}-v_p(R)\right]\,,
\end{eqnarray}

\noindent
where 
$p_T$ is the thermal pressure inside the outflow, 
$p_B$ is the magnetic pressure, 
$p_x$ is the pressure of the external IGM, and 
$M_d(R)$ is the mass of matter lying inside the shell. 
In equation~(\ref{eq-late-R}), 
the first term represents the pressure gradient driving the outflow outwards. 
The second term is the gravitational deceleration caused by 
matter existing inside the outflow radius 
(attraction between the outflow shell and the background halo + host galaxy), 
and the self gravity of the shell. 
The third term is an acceleration due to the cosmological constant. 
The final term is a drag force caused by sweeping up the IGM and 
accelerating it 
from velocity $v_p$ to $\dot{R}$. 

This formulation for the evolution of outflow radius 
[eq.~(\ref{eq-late-R})] is almost same as that of FL01, 
except that we have a factor of $[1-\cos (\alpha/2)]$ in the first term 
to take into account of the anisotropic shape of the outflow. 
\citet{tegmark93} and PMG07 used a similar equation, but 
they do not have the following terms which we include for AGN outflows: 
the magnetic pressure ($p_B$), 
the deceleration due to self-gravity of the shell ($GM_s/2R^2$), 
and the acceleration due to the cosmological constant ($\Omega_\Lambda H^2R$). 


The external gas pressure is obtained using $p_x(z) = \rho_x(z) k T_x / \mu$, 
where the external temperature is fixed at $T_x = 10^4$ K assuming 
a photoheated ambient medium, and $\mu = 0.611$ amu is the mean molecular 
mass, assuming that the medium ahead of the outflow has been photoionized
(PMG07). The working expression for the pressure of the IGM is then 
\begin{equation} 
p_x(z) = \Omega_{b,0} \frac{3 H_0^2 k T_x}{8 \pi G \mu} (1+z)^3. 
\end{equation}

Making again the approximation 
that the density of the background through which 
the AGN outflows propagate is equal to the mean matter density of the universe,
we obtain the halo mass enclosed within the shell,
\begin{equation} 
\label{Md}
M_d(R,z)=\frac{4 \pi R^3}{3}\bar\rho(z)=\frac{\Omega(z) H^2(z) R^3}{2G}. 
\end{equation} 

The thermal pressure is provided by the jet kinetic power when the AGN is 
active. It undergoes expansion losses and varies at a rate 
\begin{equation} 
\dot p^{\phantom1}_T
=\frac{\Lambda}{2\pi R^3\left[1-\cos\left(\alpha/2\right)\right]} 
-5p^{\phantom1}_T\frac{\dot{R}}{R}\,. 
\end{equation}

\noindent
The first and second terms represent the increase in pressure caused by 
injection of thermal energy into the outflow,
and the drop caused by the $pdV$ work done as the outflow expands,
respectively. The total luminosity, $\Lambda$, is the combined rate of 
energy deposition and dissipation within the outflow,
\begin{equation} 
\label{eq-L-total}
\Lambda=L_{\rm AGN}-L_{\rm Comp}-L_{\rm ne^2}-L_{\rm ion}+L_{\rm diss}\,,
\end{equation} 

\noindent
where $L_{\rm AGN}$ is the luminosity of the AGN 
contributing to the expansion of the outflow, 
$L_{\rm Comp}$ is the inverse Compton cooling off the CMB photons, 
$L_{\rm ne^2}$ is the cooling due to two-body interactions, 
$L_{\rm ion}$ is the cooling due to ionization, and 
$L_{\rm diss}$ is the heat dissipated from collisions between the expanding 
shell and the IGM. We assume that the first two terms, AGN luminosity and 
inverse Compton cooling, dominates (following PMG07) and neglect the other 
terms. The Compton luminosity is given by
\begin{equation}
\label{eq-Lcomp}
L_{\rm Comp}={2\pi^3\over45}\left({\sigma^{\phantom1}_T\hbar\over m_e}\right)
\left({kT_{\gamma0}\over\hbar c}\right)^4\left(1-\cos{\alpha\over2}\right)
(1+z)^4p^{\phantom1}_TR^3\,,
\end{equation}

\noindent (PMG07) where $T_{\gamma0}$ is the temperature of the CMB at
present. 

Magnetic fields thread AGN jets from sub-pc to kpc scales, 
but their detailed characteristics are not well known. 
Following FL01, we make the simplifying assumption that during its activity 
period an AGN ejects magnetic energy equal to a fixed fraction of the 
kinetic energy injected by the jets, $E_B = \epsilon_B E_K$. 
The fraction $\epsilon_B$ is taken as a free parameter, 
with a canonical value of $\epsilon_B = 0.1$. 
We also assume that the magnetic field (of strength $B_m$) is tangled, 
and exerts an isotropic pressure $p^{\phantom1}_B=(1/3)B_m^2/8\pi$ 
affecting the dynamics of the outflow.
The relation between energy, pressure and volume is
$E_B=3p^{\phantom1}_BV$. 
From this, we can derive the
equation for the evolution of the magnetic pressure,
\begin{equation} 
\label{pmag}
\dot p^{\phantom1}_B
=\frac{\epsilon^{\phantom1}_BL_{\rm AGN}}
{4\pi R^3\left[1-\cos\left(\alpha/2\right)\right]} 
-4p^{\phantom1}_B\frac{\dot{R}}{R}\,. 
\end{equation}

\noindent where the last term comes from the conservation
of magnetic flux of a magnetic field frozen into the expanding outflow.
At early times when the AGN is active, the first term in the right-hand side dominates,
and we get $p^{\phantom1}_B/p^{\phantom1}_T = \epsilon^{\phantom1}_B/2$. 
After the AGN turns off, we
set $L_{\rm AGN}=0$ (see below), and equation~(\ref{pmag}) gives
$p^{\phantom1}_B\propto R^{-4}$. 
In our simulations we find that the thermal pressure always 
dominates over the magnetic pressure of the outflows. 

The combination of equations~(\ref{lagn})--(\ref{pmag}) fully
describe the evolution of the outflows. In Appendix A, we describe how
these equations are solved in practice.


In our simulations we allow each AGN to evolve through an active life 
when $z_{\rm bir} > z > z_{\rm off}$. 
For this time period $T_{\rm AGN}$, we use the jet kinetic luminosity 
as the AGN luminosity in equations~(\ref{eq-L-total}) and~(\ref{pmag}), 
$L_{\rm AGN} = L_K = \epsilon^{\phantom1}_K L_{\rm bol}$. 

After the central engine has stopped activity (when $z < z_{\rm off}$), 
it enters the post-AGN dormant phase. Then we set the AGN luminosity 
to zero ($L_{\rm AGN} = 0$) in equations~(\ref{eq-L-total}) and~(\ref{pmag}).
The gas inside the outflow is still overpressured relative to
the IGM, so the expansion continues
(e.g., \citealt{kronberg01, reynolds02}, PMG07, \citealt{barai08}),
but the pressure drops faster since there is no energy input from the AGN. 
The outflow keeps expanding as long as its pressure 
$p^{\phantom1}_T+p^{\phantom1}_B$ exceeds 
the external pressure $p_x$ of the IGM.
When $p^{\phantom1}_T+p^{\phantom1}_B=p_x$, 
the outflow has reached pressure equilibrium.
After this point, the outflow simply evolves passively with the Hubble flow. 


\subsection{Metal Enrichment}
\label{sec-num-metal} 


The material transported into the IGM by AGN outflows, for the
most part, does not originate from the AGN itself, but from the
interstellar medium of the host galaxy. Stellar evolution in the host
galaxy have enriched the interstellar medium with heavy
elements, or metals, and these metals will be carried into the
surrounding IGM by the outflows. Using our simulations, we can 
calculate the distribution and concentration of metals in the IGM.
Here we focus on the distribution of metals and the
{\it enriched volume fraction}, that is, the fraction of
the IGM by volume that has been enriched by AGN outflows. The concentration
of metals in the IGM will be addressed in a forthcoming paper. 

To calculate the fraction of the total volume filled by outflows,
we cannot simply add up the final volumes of the outflows,
and divide by the volume of the computational box. Intergalactic
gas enriched by outflows will move with time as structures grow, and
therefore regions that were never hit by outflows might end up
containing metals. 
We take this effect into account,
by employing a dynamic particle enrichment scheme to quantify 
the cosmological enrichment history of the IGM. 
During the evolution of an outflow, we identify the particles that the
outflow hits. These particles are flagged as been enriched. 
This is done at every timestep for every outflow 
present in the simulation box at that time.
The fraction of the
volume of the box occupied by these enriched particles is 
then estimated, as follows. 

We divide the computational
volume into $N_{\rm ff}^3 = 256^3$ cubic cells, and identify which
cells contain matter that has been enriched by outflows. Notice that
we cannot simply count the number of cells that contained particles
that have been flagged as enriched. This approach works in high-density
regions, but fails in low-density regions where the cells are smaller
than the local particle spacing. We solve this problem by using
a Smoothed Particle Hydrodynamics technique. 
A smoothing length $h$ is ascribed to each particle. 
We calculate $h$ iteratively by requiring that each particle 
has between $60$ and $100$ neighbors within a distance $1.7h$.
We then treat each particle as an extended sphere of radius $1.7h$ 
over which it is considered to be spread. 
The cells that are covered by one or more enriched particles are then
considered enriched. This method works well both in low- and
high-density regions.
The total number of filled cells, $N_{\rm rich}$, gives the total volume 
of the box occupied by the enriched particles. 
The fractional volume of the simulation box enriched by AGN outflows 
is then $N_{\rm rich}/N_{\rm ff}^3$. Notice that this is not done during
the simulation itself. The simulation produces dumps at various redshifts
containing the positions and velocities of particles, as well as the
flags which indicate which particles are enriched in metals. The
calculation of the enriched volume fraction is then 
performed as post-processing.


\section{RESULTS AND DISCUSSION} 
\label{sec-results}

\begin{deluxetable*}{cccccc}
\tabletypesize{\scriptsize}
\tablecaption{Parameters of the Simulation 
and Final Enriched Volume Fractions}
\tablewidth{0pt}
\tablehead{ 
\colhead{Run} & 
\colhead{$\alpha$ ($^{\circ}$)} & 
\colhead{$T_{\rm AGN}$ (yr)} & 
\colhead{$\epsilon_K$} & 
\colhead{Bias in Location} & 
\colhead{$N_{\rm rich}/N_{\rm ff}^3 (z=0)$} 
}
\startdata 
A & $180$ & $10^8$ & $0.10$ & $\times$ & 0.80 \cr
B & $120$ & $10^8$ & $0.10$ & $\times$ & 0.82 \cr
C & $60$ & $10^8$ & $0.10$ & $\times$ & 0.83 \cr 
\noalign{\hrule} 
D & $60$ & $10^7$ & $0.10$ & $\times$ & 1.00 \cr
E & $60$ & $10^9$ & $0.10$ & $\times$ & 0.75 \cr
\noalign{\hrule} 
F & $60$ & $10^8$ & $0.05$ & $\times$ & 0.79 \cr
G & $60$ & $10^8$ & $0.01$ & $\times$ & 0.71 \cr
\noalign{\hrule} 
H & $60$ & $10^8$ & $0.10$ & $\surd$ & 0.75 \cr
I & $60$ & $10^8$ & $0.10$ & $\surd$ & 0.65 \cr
\enddata
\label{TabRuns}
\end{deluxetable*} 

We performed a series 9 simulations by varying some parameters of our AGN 
evolution model. 
Table~\ref{TabRuns} summarizes the characteristics of each run. 
The first column gives the letter identifying the run. In
columns 2, 3, and 4, we list the opening angle of the outflows,
the lifetime of the AGN's, and the kinetic fraction, respectively.
The fifth column indicates if biasing was used when 
determining the locations of the AGN's. 
The results are presented and discussed in the following subsections. 
For each run we plot the fraction of the cosmological volume 
enriched by the AGN outflows. 
The last column of Table~\ref{TabRuns} lists the resulting 
metal-enriched volume fraction at the present epoch obtained for each run. 

\subsection{Evolution of a Single Outflow} 
\label{sec-single} 

\begin{figure} 
\centering
\includegraphics[width = \linewidth]{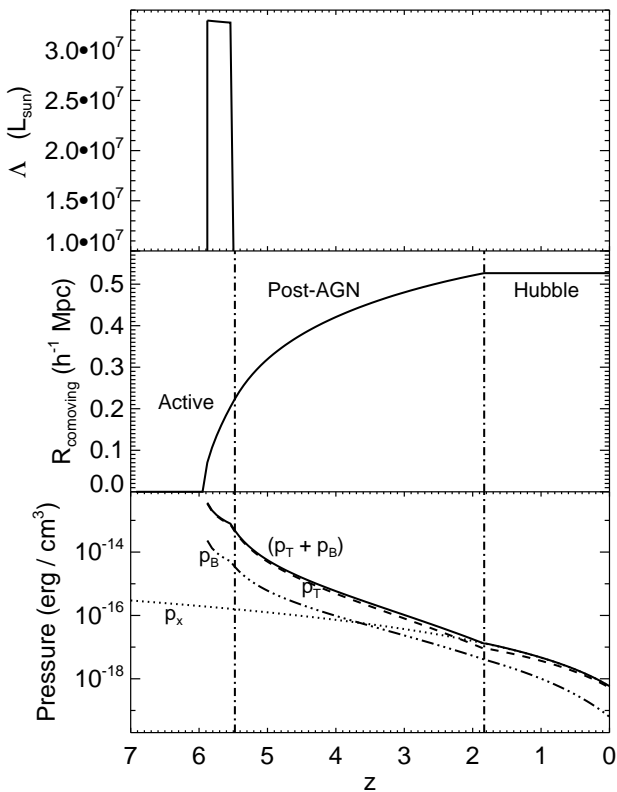}
\caption{ 
Characteristic quantities for the evolution of a single outflow with 
lifetime $T_{\rm AGN}=10^8{\rm yr}$, 
bolometric luminosity $L_{\rm bol} = 3.3 \times 10^8 L_{\odot}$, 
and opening angle $\alpha = 60^{\circ}$, as a function of redshift. 
Upper panel: total luminosity 
[eq.~(\ref{eq-L-total}), $\Lambda=L_{\rm AGN}-L_{\rm Comp}$];
middle panel: comoving radius of outflow ($R_{\rm comoving}$);
lower panel: 
thermal pressure inside the outflow ($p_T$, {\it dashed}), 
magnetic pressure ($p_B$, {\it dash-dot-dot-dot}), 
total outflow pressure ($p_T+p_B$, {\it solid}), 
and pressure of the external IGM ($p_x$, {\it dotted}). 
The vertical {\it dash-dotted lines} in the lower two panels 
separate the different phases of expansion of the outflow: 
active, post-AGN overpressured, and the final passive Hubble evolution. 
}
\label{fig-single}
\end{figure}

Figure~\ref{fig-single} shows the redshift evolution of a single anisotropic
outflow with opening angle
$\alpha = 60^{\circ}$, born at $z \simeq 5.9$. 
The different phases of the outflow expansion are separated by vertical lines: 
active phase of the AGN on the left, post-AGN phase in the middle, 
and passive Hubble expansion on the right. 
The top panel shows the total luminosity 
($\Lambda = L_{\rm AGN}-L_{\rm Comp}$), 
the middle panel shows the comoving radius of the outflow, 
and the bottom panel shows the pressures associated with the outflow. 

Soon after birth, the outflow is driven by the active AGN, 
whose luminosity dominates over the Compton luminosity, and it 
grows rapidly in size. 
At the end of the active phase ($z \simeq 5.5$), the AGN turns off 
and the only contribution to the total luminosity is the energy 
loss by Compton drag. 
The interior of the outflow is still overpressured 
with respect to the external IGM by a factor 
$(p^{\phantom1}_T + p^{\phantom1}_B) / p_x \sim 400$. 
So it continues to expand while its pressure falls faster 
(notice the change of slope in the bottom panel).
Finally, when the total outflow pressure falls to the level of 
the external pressure, the expansion stops.
From $z \simeq 1.8$ the comoving radius of the outflow remains constant at 
$0.53 h^{-1}{\rm Mpc}$ in the passive Hubble flow phase.  



\subsection{Opening Angle of Anisotropic Outflows} 
\label{sec-alpha} 

\begin{figure} 
\centering
\includegraphics[width = \linewidth]{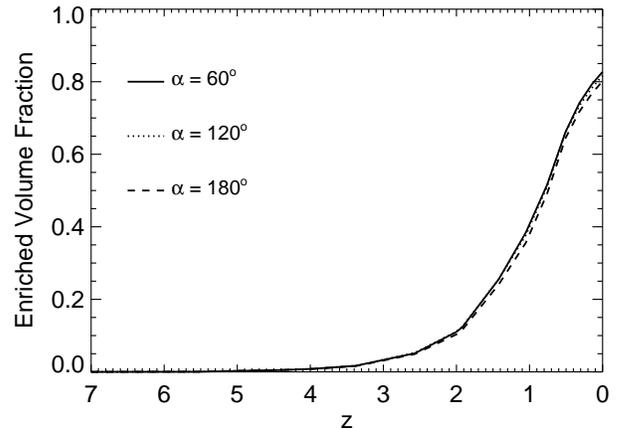}
\caption{ 
Fractional volume ($N_{\rm rich} / N_{\rm ff}^3$) of simulation box 
enriched by AGN outflows as a function of redshift, for various
opening angles: $\alpha = 60^{\circ}$ ({\it solid}), 
$120^{\circ}$, ({\it dotted}), $180^{\circ}$ ({\it dashed}). 
See \S\ref{sec-alpha} for details. 
} 
\label{fig-volFill-alpha}
\end{figure} 

To study the effect of varying opening angle of the outflows 
on the enriched volume fraction, we performed runs A, B, and C with
opening angles 
$\alpha = 180^{\circ}, 120^{\circ}$, and $60^{\circ}$, respectively. 
Here $\alpha = 180^{\circ}$ corresponds to the isotropic outflow case 
(the geometry used by FL01, LG05, and other previous studies), 
and $\alpha = 60^{\circ}$ represents the most anisotropic outflow
we consider. 

Figure~\ref{fig-volFill-alpha} shows the redshift evolution of the 
enriched volume fractions we obtained for different opening angles. 
The enriched volume fractions are small at $z > 3$, reach $0.1$ at $z \sim 2$, 
and grow rapidly afterward to $0.4$ at $z \sim 1$. 
At the present epoch, $z=0$, 80\% 
of the universe is enriched by AGN outflows. 
We do not find much dependence on the opening angle. 
More anisotropic outflows (smaller $\alpha$) are found to enrich slightly 
larger volumes. 
Such a trend is counter-intuitive since, for a given radius,
an outflow with $\alpha = 180^{\circ}$ occupies a larger volume 
than one with $\alpha = 60^{\circ}$. 
But according to our model prescription in 
\S\ref{sec-num-outflow}, 
an outflow with a smaller $\alpha$ grows to a larger radius 
than a more isotropic one, because the energy is concentrated into
a smaller volume, resulting in a larger pressure. 
Still, isotropic outflows tend to have
larger individual volumes than anisotropic ones, as PMG07 showed.

This effect is compensated by another effect: overlapping outflows.
Because we locate AGNs in dense regions, they are strongly
clustered, especially at low redshift. A dense region
will eventually harbor many AGNs\footnote{They might be active
at different epochs, though.} and their outflows will likely
overlap. Isotropic outflows tend to overlap with one another more
than anisotropic ones, for two reasons. First, for isotropic outflows 
the volume occupied by each outflow is larger, hence 
the probability of overlap is also larger. 
Second, anisotropic outflows propagate along the direction of least resistance.
As PMG07 showed, if several outflows originate from a common structure, like
a cosmological filament or pancake, they tend to align themselves along
the direction normal to the structure. This greatly reduces the amount of
overlap, especially at small opening angle $\alpha$.
In our simulations with larger $\alpha$, the outflows overlap more,
resulting in a smaller enriched volume fraction. 
So the combined effect of smaller individual volumes and less overlap cause 
the more anisotropic outflows to enrich a slightly larger volume fraction
of the IGM. 


Figures~\ref{fig-sliceA} and~\ref{fig-sliceC} 
show the evolution of the large-scale structure 
and the distribution of metals in a slice of 
comoving thickness $2h^{-1}\rm{ Mpc}$, for isotropic outflows
(run A), and outflows with opening angle of $60^\circ$
(run C). At $z=3.38$, the metals are just 
starting to emerge from the dense regions that are hosting AGNs. As
the simulation progresses, new AGNs are formed, producing additional
outflows, while the ones already present keep expanding until they
reach pressure equilibrium with the external IGM. By $z=1.05$, a significant
fraction of the volume is enriched, but metals are still concentrated near
dense regions. By $z=0$, the metals have spread into the low-density 
regions, and only the very-low-density regions, located far from any 
cosmological structure, have not been enriched.

There is no striking difference between the two figures, though we can notice
that voids are more enriched at $z=1.05$ in run C than in run A.
To illustrate this more clearly, we plot in Figure~\ref{fig-sliceAC}
the difference between the two enrichment maps. Red indicates regions that 
are enriched in run A but not in run C, while green indicates regions
enriched in run C but not in run A. Since the enriched volume fraction 
is about the same
for both runs, we expect comparable amounts of red and green, which is
indeed the case. However, their distributions are significantly different.
Looking at redshift $z=1.05$ (bottom-left panel of Fig.~\ref{fig-sliceAC}),
we find green areas all over the slice. Colored regions found inside
deep voids are almost exclusively green, while red regions are predominantly
located inside or near dense regions. Overall, anisotopic outflows enrich
low-density regions more than isotropic outflows, at the expense of
high density regions. This effect is seen most clearly at the
intermediate redshift $z=1.05$. At earlier times, the enriched volume 
fraction is
small in both simulations, making the difference small, while at late times 
most regions are enriched in both runs, as the enriched
volume fraction exceeds 0.80.
Figure~\ref{fig-slice_zoom} shows a zoom-in of the upper right region, which
illustrates the distribution of red and green areas relative 
to dense structures.

\begin{figure} 
\hskip-0.7in
\caption{ 
Slice (of comoving size $128~h^{-1}$ Mpc $\times$ $128~h^{-1}$ Mpc, 
and $2h^{-1}\rm{ Mpc}$ comoving thickness) 
off the computational volume for run A, 
showing the evolution of the distribution of metals. The areas shown in orange
are enriched. The black dots represent the PM particles, and show the
large-scale structures.
} 
\label{fig-sliceA}
\end{figure}

\begin{figure}
\hskip-0.7in
\caption{Same as Figure~\ref{fig-sliceA}, for run C.}
\label{fig-sliceC}
\end{figure}

\begin{figure}
\hskip-0.7in
\caption{ 
Differential enrichment map between runs A and C, calculated from
Figures~\ref{fig-sliceA} and~\ref{fig-sliceC}. Red areas show regions
enriched in Run A but not in Run C; green areas show regions
enriched in Run C but not in Run A.}
\label{fig-sliceAC}
\end{figure}

\begin{figure}
\hskip-0.7in
\includegraphics[width = \linewidth]{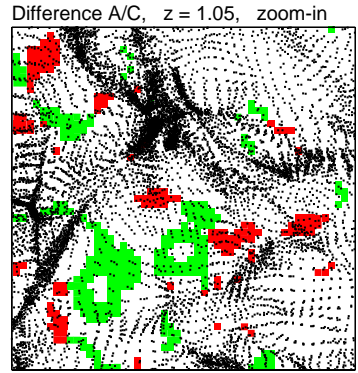}
\caption{ 
Differential enrichment map between runs A and C, at redshift $z=1.05$.
This figure shows a zoom-in of the upper right region of the third
panel of figure~\ref{fig-sliceAC}.}
\label{fig-slice_zoom}
\end{figure}

\subsubsection{Enrichment of Overdense vs.\ Underdense IGM} 

\begin{figure} 
\centering
\includegraphics[width = \linewidth]{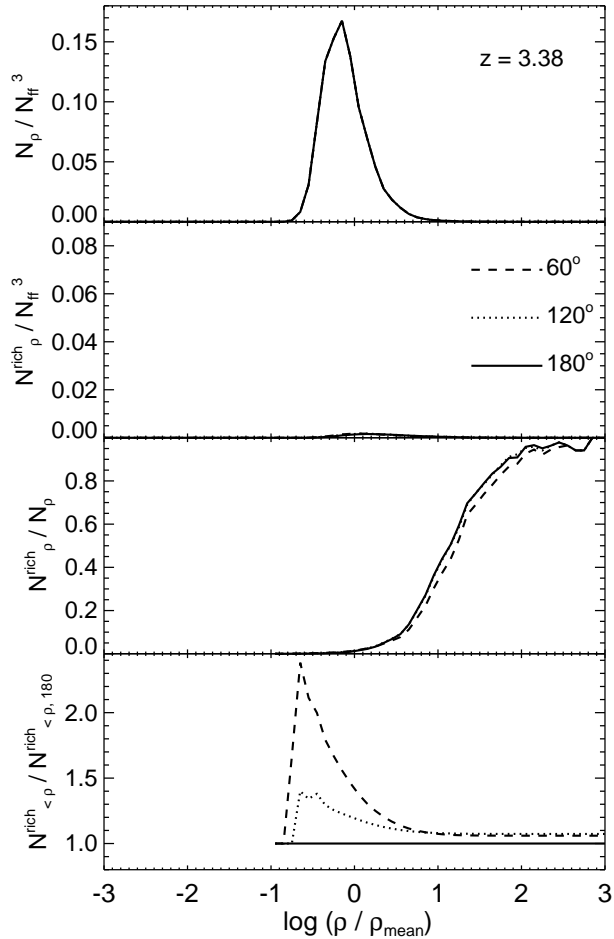}
\caption{ 
Density statistics at redshift $z = 3.38$ for runs A, B, and C.
The quantities $N_{\rho}$, $N^{\rm rich}_\rho$,
$N^{\rm rich}_{< \rho}$, and $N_{\rm ff}^3=256^3$ designate
the number of cells (whether enriched or not) at a given density,
the number of cells enriched by AGN outflows at a given density,
the number of cells enriched by AGN outflows below a given density
threshold, and the total number of cells in the computational volume,
respectively.
Top panel: histogram of the density distribution of cells. 
Second panel: number of cells enriched by AGN outflows at a given
density, divided by the
total number of cells in the computational volume.
Third: number of cells enriched by AGN outflows at a given
density, divided by the
total number of cells {\it at that density}.
Bottom: number of cells enriched by AGN outflows 
below a certain density threshold, divided
by the corresponding number for the isotropic 
case ($\alpha=180^\circ$). 
The linestyles indicate the different opening angles: 
$\alpha=60^{\circ}$ ({\it dashed}), 
$120^{\circ}$ ({\it dotted}), $180^{\circ}$ ({\it solid}). 
} 
\label{fig-fillStat-z3.38} 
\end{figure}

\begin{figure} 
\centering
\includegraphics[width = \linewidth]{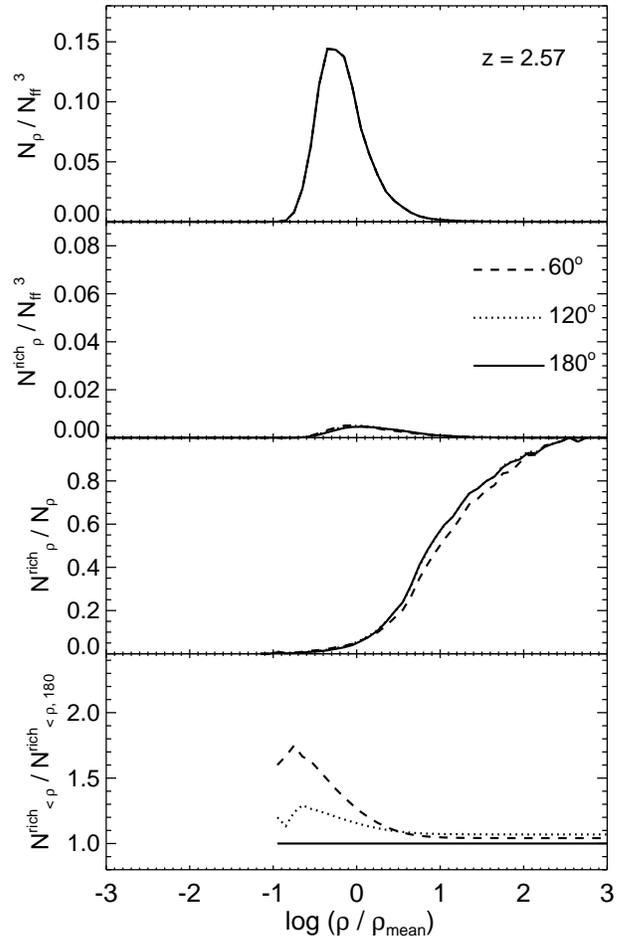}
\caption{ 
Same as Figure~\ref{fig-fillStat-z3.38}, at redshift $z = 2.57$. 
} 
\label{fig-fillStat-z2.57}
\end{figure}

\begin{figure} 
\centering
\includegraphics[width = \linewidth]{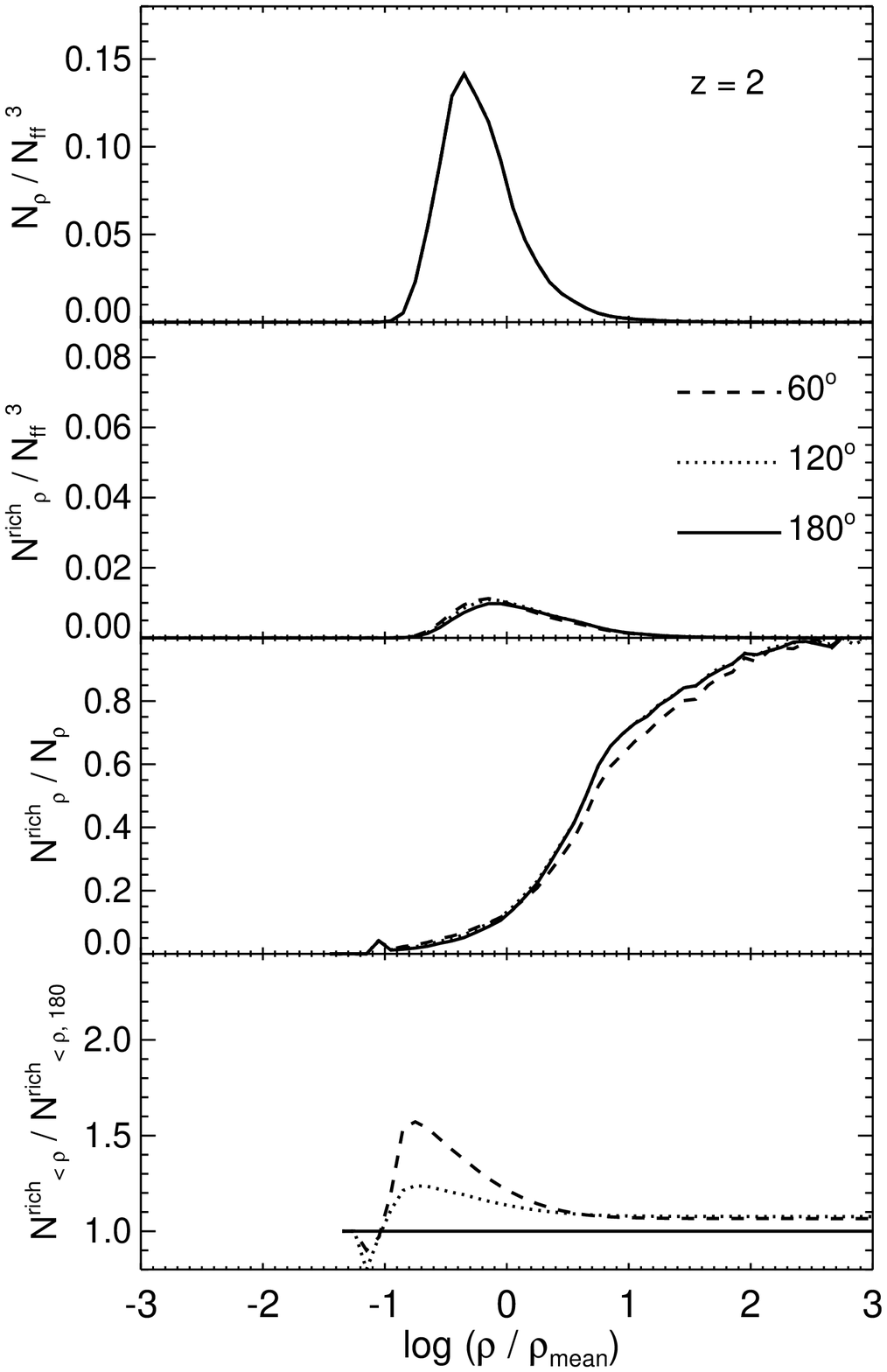}
\caption{ 
Same as Figure~\ref{fig-fillStat-z3.38}-\ref{fig-fillStat-z2.57},
at redshift $z = 2.00$. 
} 
\label{fig-fillStat-z2}
\end{figure}

\begin{figure} 
\centering
\includegraphics[width = \linewidth]{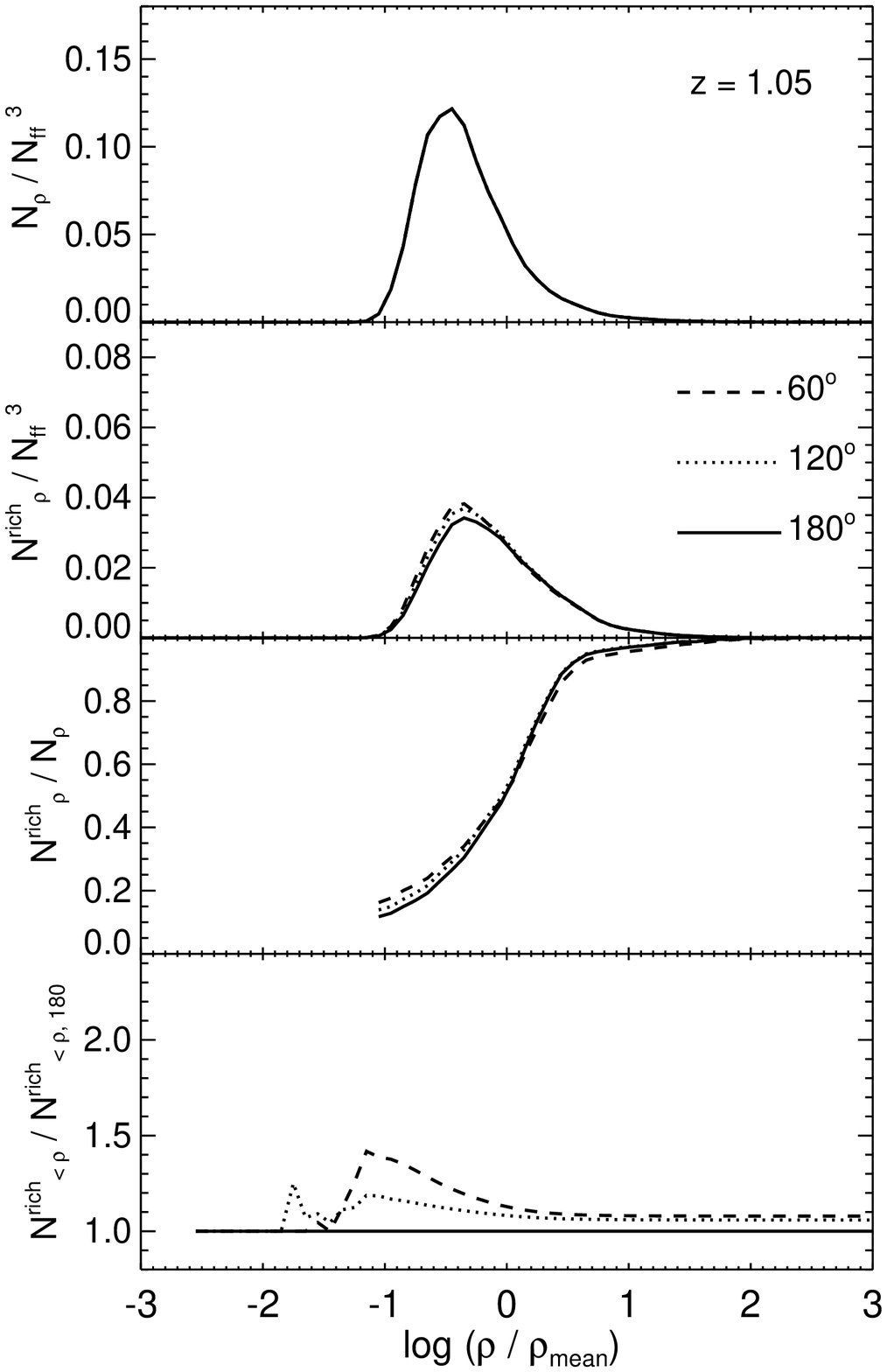}
\caption{ 
Same as Figure~\ref{fig-fillStat-z3.38}-\ref{fig-fillStat-z2}, 
at redshift $z = 1.05$. 
} 
\label{fig-fillStat-z1.05}
\end{figure}

\begin{figure} 
\centering
\includegraphics[width = \linewidth]{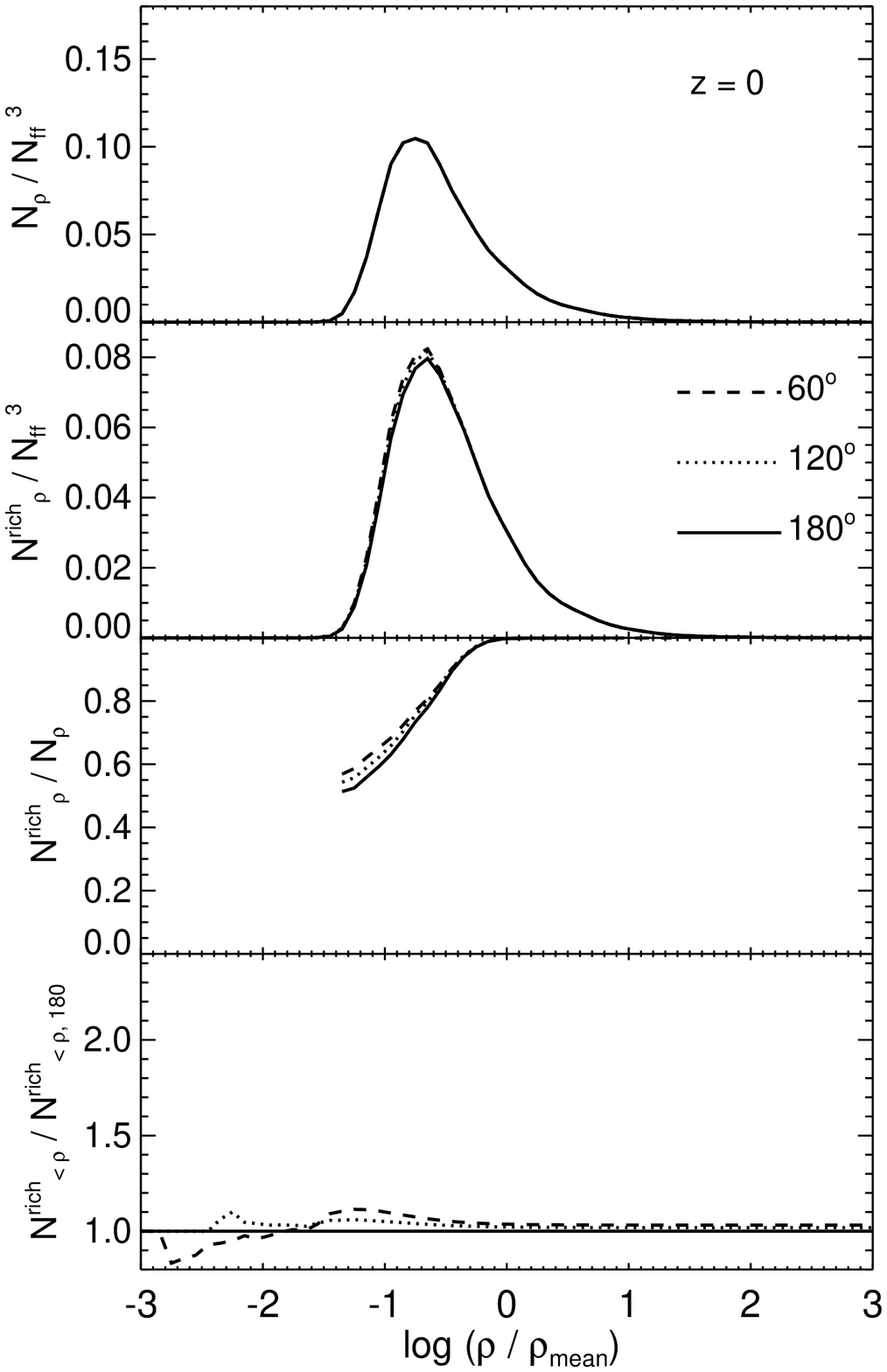}
\caption{ 
Same as Figure~\ref{fig-fillStat-z3.38}-\ref{fig-fillStat-z1.05}, 
at redshift $z = 0.00$. 
} 
\label{fig-fillStat-z0}
\end{figure}

We investigate the relationship between metal enrichment 
by AGN outflows and the
density of the regions being enriched, using runs A, B, and~C. 
Figures~\ref{fig-fillStat-z3.38} -- \ref{fig-fillStat-z0} 
show various density statistics of the enriched
regions at five different redshifts,
$z = 3.38, 2.57, 2.0, 1.05$ and $0$,
for 3 opening angles, $\alpha = 60^{\circ}, 120^{\circ}$, and $180^{\circ}$.
All quantities are plotted as functions of the overdensity 
(ratio of the local density to the mean density, $\rho / \rho_{\rm mean}$).

The top panels show the fractional number of cells in the 
whole grid 
(whether enriched or not) at a density $\rho$. 
The distribution resembles a skewed Gaussian, 
with more underdense volumes than overdense. 
The Gaussian becomes wider in time as large scale structures form in the 
cosmological volume, with further higher and lower densities 
being reached gradually. 
Most of the volume of the simulation box
is underdense, 
and the peak of the distribution shifts to lower densities with time. 

The second panels show the fractional number of cells
which are enriched by AGN outflows ($N^{\rm rich}_{\rho}$). 
The enriched volume fraction is very small at early time, 
and increases with time, becoming significant from $z \sim 2-3$ and 
afterward. 
At a given redshift, 
we see a slight trend of the peak of the resulting Gaussian distribution 
shifted to lower densities for increasing anisotropy 
($\alpha = 180^{\circ}, 120^{\circ}, 60^{\circ}$). 
This indicates that more anisotropic outflows enrich underdense regions of 
the IGM. This is most noticeable at redshift~$z=1.05$
(Fig.~\ref{fig-fillStat-z1.05}).

The third panels show the ratio of the cell counts in the second to the 
top panel, 
i.e., the number of cells enriched as the fraction of the total number of 
cells. 
In high-density regions ($\rho>\rho_{\rm mean}$) 
more isotropic outflow enrich larger volumes
than anisotropic ones, while in
low-density regions ($\rho<\rho_{\rm mean}$) we find
exactly the opposite behavior.
The effect is small, but most prominent at $z = 1.05$. At $z=0$, dense 
regions are fully enriched for all values of $\alpha$, while at $z>2$
low-density regions are just starting to get enriched.

The bottom panels are cumulative versions of the second panels, 
showing the number of enriched cells below a given density 
threshold ($N^{\rm rich}_{< \rho}$), divided by the
corresponding number $N^{\rm rich}_{< \rho,180}$ for
isotropic outflows.\footnote{Note that $N^{\rm rich}_{< \rho}=N^{\rm rich}$
in the limit of large $\rho$. Hence, the fact that the curves reach different
asymptotes in the right hand side of the panels simply reflects the
weak dependence of the overall enriched volume fraction on opening
angle.}. In low-density regions [$\log(\rho/\rho_{\rm mean})<0$],
the ratio $N^{\rm rich}_{< \rho}/N^{\rm rich}_{< \rho,180}$
strongly depends on the opening angle, at all redshift but
especially at early time ($z\geq2$).
The most anisotropic outflow ($\alpha = 60^{\circ}$) enriches 
the largest underdense volumes, and the effect is stronger 
at lower densities.
The effect is most prominent at earlier epochs ($z \sim 3.5 - 2$), 
and decreases in intensity with time, but is
still visible at the present epoch. 
We explain this trend as follows.
At earlier times, when the AGN outflows have just started to enrich the 
surrounding volumes, 
more anisotropic outflows enrich larger fractions of the lower density IGM, 
since these outflows expand along the direction of least resistance. 
As time goes on, large-scale structures form and grow, causing material to 
accrete from low- to high-density regions. 
This tends to even out the density of the enriched volumes, and as time
goes on the effect of preferential enrichment of low-density regions 
by anisotropic outflows gets washed out. 
Such a trend in seen in our resulting density statistics, 
where the difference between the different opening angles is very small in 
Figure~\ref{fig-fillStat-z0} at $z=0$. 

\subsubsection{Comparison with Pieri et al. 2007} 

The bottom 3 panels of
Figures~\ref{fig-fillStat-z3.38}--\ref{fig-fillStat-z0} 
are analogous to the plots in Figure~9 of PMG07.
These authors considered outflows driven by SNe explosions in dwarf
galaxies, and found that the anisotropy of the outflows had
a major impact on the IGM enrichment in low-density regions.
Outflows with opening angle $\alpha = 60^{\circ}$ enriched the 
lowest-density regions of the IGM 
by a factor $3 - 4$ more than isotropic outflows. 
In this study, where we consider outflows driven by AGN 
in active galaxies, we find that a similar trend 
of increasingly anisotropic outflows to enrich lower-density regions, 
but the effect
depends significantly on the redshift. 
The trend we obtained is not as dramatic as in PMG07
at the present epoch (Fig.~\ref{fig-fillStat-z0}),
but becomes more and more prominent at earlier epochs. 
We find that at $z = 2.57$ (Fig.~\ref{fig-fillStat-z2.57}), the factor by 
which outflows with $\alpha = 60^{\circ}$ enrich low-density regions 
more than isotropic ones goes up to $1.7$. 
At $z = 3.38$ (Fig.~\ref{fig-fillStat-z3.38}), the factor goes as 
high as $2.3$. 

Even though we have used the same prescription for the anisotropic outflows 
as PMG07, we point out the following differences between the two studies. 
PMG07 did not perform an actual cosmological simulation. Instead, they
generated an initial gaussian random field at high redshift, filtered that
density field at various mass scales, and used the spherical collapse model
to predict the collapse redshift of each halo
(see \citealt{sb01} and PMG07 for details). This semi-analytical model
has the advantage of simplicity, but does not include an accurate treatment
of halo mergers and accretion of matter onto halo, and furthermore it
underestimates the level of clustering of the halos
(see, however, \citealt{ppm09}).
In this paper, we argue that the overlap between outflows, which
results from the clustering of halos, is more important for isotropic
outflows than anisotropic ones, and that accretion of low-density,
metal-enriched matter onto dense structures tends to erase the effect
of anisotropy. As a consequence, we find that the effect of anisotropy
is still important at high redshifts, but greatly reduced at low redshifts.

\subsection{AGN Lifetime} 
\label{sec-Tagn} 

\begin{figure} 
\centering
\includegraphics[width = \linewidth]{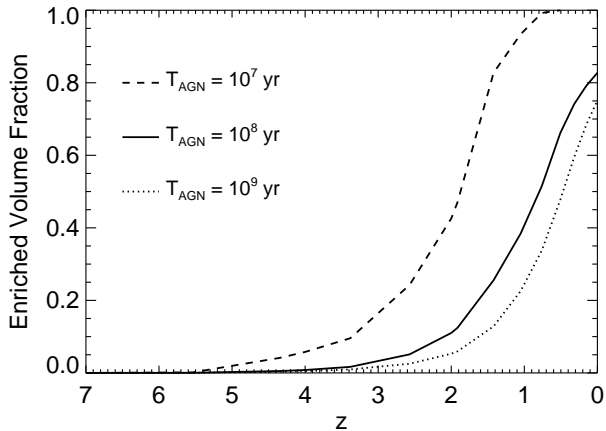}
\caption{ 
Fractional volume ($N_{\rm rich} / N_{\rm ff}^3$) of simulation box 
enriched by AGN outflows as a function of redshift,
for different AGN activity lifetimes: 
$T_{\rm AGN} = 10^7{\rm yr}$ ({\it dashed}), $10^8{\rm yr}$ ({\it solid}), 
$10^9{\rm yr}$ ({\it dotted}). 
See \S\ref{sec-Tagn} for details. 
} 
\label{fig-volFill-TAGN}
\end{figure} 

\begin{deluxetable}{cccc}
\tablecaption{Runs with different active lifetimes $T_{\rm AGN}$}
\tablewidth{0pt}
\tablehead{ 
\colhead{Run} & 
\colhead{$T_{\rm AGN}$ (yr)} & 
\colhead{Population Size} & 
\colhead{$N_{\rm rich}/N_{\rm ff}^3\;(z=0)$} 
}
\startdata 
D & $10^7$ & $15\,279\,029$ & 1.00 \cr
C & $10^8$ & $1\,535\,362$ & 0.83 \cr 
E & $10^9$ & $162\,228$ & 0.75 \cr
\enddata
\label{TabTagn}
\end{deluxetable} 

The simulations A, B, and C presented in the previous section all
assumed that AGN activity lasts $T_{\rm AGN} =10^8{\rm yr}$. To study
the effect of varying $T_{\rm AGN}$, we performed runs D and E,
using AGN activity lifetimes of $T_{\rm AGN}=10^7{\rm yr}$ and $10^9{\rm yr}$,
respectively. These values correspond to the lower limit and
upper limit on $T_{\rm AGN}$ used by LG05.
Simulations C, D, and E use the same opening angles, $\alpha = 60^{\circ}$,
and the same kinetic fraction, $\epsilon^{\phantom1}_K = 0.1$,
and differ only in the value of $T_{\rm AGN}$.
We neglect any repeated activity (which might happen because of duty cycle) 
of the same AGN after it has become inactive, as did LG05. 
Considering duty cycles would have a small effect in our results, 
because of the way we model the outflow expansion and evolution 
(\S\ref{sec-num-outflow}, our outflows continue to remain in the 
simulation box even after the central AGN has died). 

A value of few $\times 10^8{\rm yr}$ is considered as the typical 
activity lifetime of the 
SMBH at the centers of active galaxies in other studies \citep[e.g.,][]{yu08}. 
A lifetime of $5 \times 10^8{\rm yrs}$ has been used in 
theoretical modeling of radio galaxies by \citet{BRW, barai07}. 
Observations of X-ray activity in AGN \citep{barger01}, 
SDSS optical studies of active galaxies \citep{miller03},
and black hole demographics arguments \citep[e.g.,][]{marconi04} 
all support an AGN activity lifetime of $\sim 10^8{\rm yr}$ or more 
\citep[also,][]{mclure04}. 

Table~\ref{TabTagn} gives the parameters and results of runs C, D and E. 
The first and second columns give the run name and the value of 
$T_{\rm AGN}$ used, respectively. 
Column 3 gives the size of the AGN population, that is, the total number of 
sources 
obtained from the QLF (\S\ref{sec-num-QLF}) in our cosmological volume 
over the Hubble time. The last column lists the resulting 
metal-enriched volume fractions at the present epoch. 
There are $\sim 10$ times more AGNs generated for a 
decrease of $T_{\rm AGN}$ by a factor of 10, since the QLF has to 
remain constant. This change in the number of sources makes a 
large difference in the resulting enriched volume fractions. 

The redshift evolution of the enriched volume fractions for 
different active lifetimes are plotted in Figure~\ref{fig-volFill-TAGN}. 
The largest AGN population ($T_{\rm AGN} =10^7{\rm yr}$) enriches 
significantly larger volume than the other two populations. 
In this case (run D) the fractional volume starts to become 
appreciable from $z \sim 3.5$, and grows afterward, 
such that $100 \%$ of the volume of the box is enriched by $z \simeq 0.7$. 
The other two populations, with $T_{\rm AGN} =10^8$ and $10^9{\rm yr}$, 
enrich $0.83$ and $0.75$ of the volume at $z = 0$, respectively. 
This result, that AGNs with shorter lifetimes enrich larger 
volumes than those with longer lifetimes, 
is opposite to what LG05 and \citet{barai08} obtained. 
We explain this in the following, stressing the point that 
the method we employed to enrich our cosmological volumes is quite 
different from 
what other studies have used.

We use a dynamic particle enrichment scheme (\S\ref{sec-num-metal}), 
whereby we marked particles lying inside outflow volumes as enriched, 
assign a smoothing length to each of them, 
and calculate the volume of the box being filled by these enriched particles. 
In other studies (LG05 and \citealt{barai08}) the enriched volume fraction
is calculated by finding the fraction of the computational volume
which is located inside the geometrical boundaries of the outflows.  
In our approach, particles that are metal-enriched by outflows 
are allowed to move,
and might move to regions not intercepted by an outflow. 
This increases the chance of enriching a larger fraction of the total volume. 
The trend heightens when the number of outflows increase; as more AGN are born 
they are placed in distinct locations distributed throughout the box 
(according to \S\ref{sec-num-locate}), 
and they could potentially enrich larger volumes. 
This
causes the enriched volume 
fraction to increase with increasing population size, 
even though the lifetime of each outflow is shorter. 

\subsection{Kinetic Fraction} 
\label{sec-epsK}

\begin{figure} 
\centering
\includegraphics[width = \linewidth]{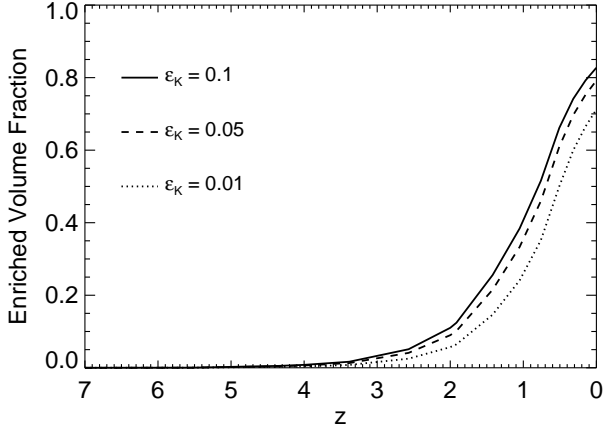}
\caption{ 
Fractional volume ($N_{\rm rich} / N_{\rm ff}^3$) of simulation box 
enriched by AGN outflows as a function of redshift, for
different kinetic fractions:
$\epsilon_K = 0.1$ ({\it solid}), 
$0.05$ ({\it dashed}), $0.01$ ({\it dotted}). 
See \S\ref{sec-epsK} for details. 
} 
\label{fig-volFill-epsK}
\end{figure} 

In runs F and G, we consider different values of the kinetic fraction 
(the fraction of the AGN bolometric luminosity which 
converts to the kinetic luminosity of the jets, \S\ref{sec-num-outflow}), 
$\epsilon_K = 0.05$ and $0.01$, 
in addition to our standard value $\epsilon^{\phantom1}_K = 0.1$ (run C).
Comparing with other studies, FL01 used a value 
$\epsilon^{\phantom1}_K=0.1$.\footnote{FL01
quote a value $\epsilon^{\phantom1}_K = 1.0$, 
but they expressed the quasar kinetic luminosity in terms of the 
B-band luminosity
instead of the bolometric luminosity.}
LG05 first experimented with a value 
$\epsilon^{\phantom1}_K = 0.1$, 
and found an enriched volume fraction that was too large
to agree with observations, so later they
chose $\epsilon^{\phantom1}_K = 0.01$ as their fiducial value. Hence,
we are sampling the range of values that have been
considered in other studies.

The resulting enriched volume fractions
are plotted in Figure~\ref{fig-volFill-epsK} as a function of redshift,
for the different kinetic fractions. We find
the same general trend as in runs A -- E, namely that the enriched 
volume fraction is small at $z>2$, 
and then grows rapidly afterward up to the present. 
At $z=0$, fractions of 0.83, 0.79, and 0.71 of the volume are enriched 
by outflows with $\epsilon^{\phantom1}_K = 0.1$, 0.05, and 0.01, respectively. 

Using their model for the AGN outflows, LG05 found that the 
cosmological volume is completely filled ($100 \%$)
by $z \sim 2$ when using $\epsilon_K = 0.1$ or $0.05$, 
and with $\epsilon_K = 0.01$, their volume filling fraction at 
$z = 0$ is $80 \%$. 
The enriched volume fractions we obtained with $\epsilon_K = 0.1$ and $0.05$ 
are comparable to those obtained by LG05 with $\epsilon_K = 0.01$. 
With $\epsilon_K = 0.01$, we obtained an enriched volume
fraction $\sim 10 \%$ smaller than that of LG05. 
We find that decreasing the kinetic fraction by a factor of 10 
reduces the volume enriched by $\approx 12 \%$. 
This is comparable to a corresponding $\approx 20 \%$ reduction 
in the volume filling fraction found by LG05 (their Fig.~3), 
when $\epsilon_K$ is reduced by a factor of 10. It may seem surprising
that increasing by a factor of 10 the energy driving the expansion
of the outflows has only a $\sim14\%$ effect on the final results.
With a larger kinetic fraction, outflows are certainly larger,
but also overlap more with one another. The lowest-density regions,
located far from any AGN, still manage to escape enrichment.

\subsection{Bias in AGN Location} 
\label{sec-bias} 

\begin{figure} 
\centering
\includegraphics[width = \linewidth]{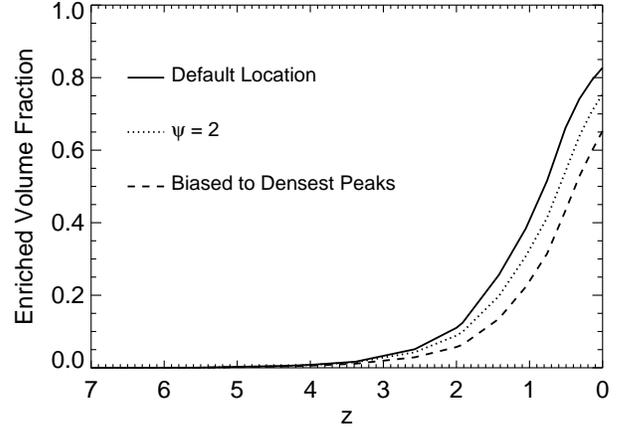}
\caption{ 
Fractional volume ($N_{\rm rich} / N_{\rm ff}^3$) of simulation box 
enriched by AGN outflows as a function of redshift,  
The linestyles indicate different methods of AGN location: 
Default run C ({\it solid}), 
run H with bias parameter $\psi = 2$, as in LG05 ({\it dotted}), 
run I, biased to densest peaks ({\it dashed}). 
See \S\ref{sec-bias} for details. 
} 
\label{fig-volFill-LocBias}
\end{figure} 

AGNs are observed to be clustered in the universe, 
with the clustering amplitude increasing with redshift 
\citep{porciani04, croom05, shen07}. 
We study the effect of bias in the distribution of AGNs 
by considering
two different methods for spatially locating the AGNs inside the
computational volume, 
in addition to our original method (\S\ref{sec-num-locate}) 
used in runs A--G.

In run H, we use a location biasing condition similar to that used by LG05, 
but restricted to cells containing a local density peak
(this is required by our method for finding the direction 
of last resistance as discussed in \S\ref{sec-num-locate}).
We calculate the probability of each peak cell of hosting an AGN as
\begin{equation} 
\label{eq-prob}
{\cal P} = 
\frac{\rho^{\psi}}{\sum_{1}^{N_{\rm peak}} \left( \rho^{\psi} \right)}\,,
\end{equation}

\noindent
where $\rho$ is the filtered density (see \S\ref{sec-num-locate}) of the cell, 
$N_{\rm peak}$ is the number of available density peaks
(not containing an AGN already) inside the computational volume, and 
$\psi$ is the bias parameter, which is denoted by $\alpha$ in equation~(2) 
of LG05. 
The summation in the denominator of equation~(\ref{eq-prob}) 
is over all the density peaks at the relevant timestep. 
We use a constant bias value $\psi = 2$, as did LG05 in their fiducial run. 
At each timestep, we use a Monte Carlo rejection
method to locate the AGNs randomly, with a probability given by
equation~(\ref{eq-prob}).

In run I, 
we bias the AGNs in favor of the highest-density peaks 
inside the cosmological volume. 
At each timestep, we find all the peak cells 
(\S\ref{sec-num-locate}), 
and then sort them in descending order of their density. 
The new AGNs (those born in that timestep) are located in the peak cells 
which have the highest densities. 
We start with the highest-density peak and locate one AGN 
(selected randomly from the ones just born) in it, 
go to the next highest-density peak cell and repeat the process, 
until all AGNs have been assigned locations. Notice that
the methods used in simulations C and I effectively correspond
to bias parameters $\psi=0$ and $\psi\rightarrow\infty$,
respectively.

Figure~\ref{fig-volFill-LocBias} shows the redshift evolution of the 
enriched volume fraction for the different 
location methods. The enriched volume fraction decreases with increasing bias,
at all redshifts. At $z=0$, the enriched volume fraction reaches
0.83 for the case without bias ($\psi=0$, run C), 0.75 for the intermediate
case ($\psi=2$, run H), and $0.65$ for the maximum bias case
($\psi\rightarrow\infty$, run I).
Such a behavior is expected, as with the introduction of gradual bias 
AGNs are increasingly  clustered in high-density regions, 
so their outflows tend to overlap. Biasing to the highest density
peaks reduces the present enriched volume fraction by 18\% compared
with the default case. 

\section{SUMMARY AND CONCLUSION} 
\label{sec-conclusion} 

We have implemented a semi-analytical model of anisotropic AGN outflows 
in cosmological N-body simulations. The AGNs are placed
at the location of local density peaks, and the outflows 
are allowed to expand anisotropically from those peaks, along the 
direction of least resistance, with a biconical geometry. 
Each AGN produces an outflow from its birth, 
which at first expands rapidly in the active-AGN phase for a time 
$T_{\rm AGN}$ until the AGN turns off, and 
then continues to expand slowly due to its overpressure. 
Finally after coming to pressure equilibrium with the external IGM the outflow 
simply follows the Hubble flow. 
The outflows carry with them metals produced by stars in the host 
galaxy, and enrich the regions of the IGM that they intercept.
We used this algorithm to
simulate the evolution of the large-scale structure and the
propagation of outflows inside 
a cosmological volume of size ($128~h^{-1}$ Mpc)$^3$, 
from initial redshift $z=24$ to final redshift $z=0$, 
in a concordance $\Lambda$CDM model.
We performed a total of
9~simulations, varying the opening angle of the outflows,
the active lifetime of the AGNs, the kinetic fraction, and the
method used for locating the AGNs inside the cosmological volume.

(1) The enriched volume fractions (fraction by volume of the IGM 
enriched in metals by the outflows) are small 
at $z \gtrsim 2.5$, and then grow rapidly afterward up to $z = 0$. 
In our simulations with different parameter values, 
we found that AGN outflows enrich $0.65 - 1.0$ of the total volume of the 
universe by the present epoch. 

(2) The enriched volume fractions do not depend significantly 
on the opening angle of the outflows. 
More anisotropic outflows (smaller $\alpha$) are found to 
enrich slightly larger volumes than more isotropic ones, 
because the anisotropic ones grow bigger in radius and have less overlap. 
Increasingly anisotropic AGN outflows enrich lower-density 
volumes of the IGM, to an extent that depends on redshift. 
The trend is most prominent at earlier epochs, 
and not very dramatic but still visible at the present. 
Outflows with opening angles $\alpha = 60^{\circ}$ 
enrich the underdense IGM more than 
isotropic outflows by a factor of $2.3$ at $z = 3.38$, and $1.7$ at 
$z = 2.57$.
Initially, high-density regions get enriched by outflows, since
the AGNs producing these outflows are located predominantly in
these regions. Eventually, expanding outflows reach low-density
regions of the IGM, and these 
regions also get enriched. This effect is larger
for more anisotropic outflows, since these outflows expand along the
directions of least resistance, reaching the low-density regions
sooner and more easily.
That enriched matter, located in underdense regions at earlier epochs, 
gravitationally accretes into higher-density regions with time, as 
large-scale structures grow. As a result,
the effect of preferential enrichment of underdense IGM by more 
anisotropic outflows gets washed out with time, 
such that the difference is quite small at the present, even though it is
quite significant at redshifts $z>1$. 

(3) Reducing the active lifetime $T_{\rm AGN}$ of the AGNs results in
larger enriched volume fractions. 
Any reduction in $T_{\rm AGN}$ must be accompanied
by a corresponding increase in the number of AGNs, such that the observed QLF
remains the same. When the number of sources increases, these sources
tend to be distributed more uniformly throughout the computational volume,
resulting in a more efficient enrichment. With $T_{\rm AGN}=10^7{\rm yr}$,
100\% of the volume is enriched by redshift $z=0.7$.

(4) The enriched volume fractions for kinetic fractions 
$\epsilon^{\phantom0}_K = 0.1, 0.05$ and $0.01$ 
are $0.83, 0.79$ and $0.71$, respectively, at the present epoch.
This is consistent with the results reported by LG05. Increasing
$\epsilon^{\phantom0}_K$ results in larger outflows, but the
main consequence is to increase the level of
overlap, rather than the enriched volume fraction.

(5) We varied the prescription for locating AGNs in the computational
volume by introducing a bias parameter which favors high-density peaks.
This increases the level of clustering of the AGNs, and when
AGNs are more clustered, there is more overlap between outflows,
resulting in a lower enriched volume fraction at all redshifts. The
maximum effect is an 18\% decrease at $z=0$ when going from an
unbiased distribution to a maximally biased one.

This paper focused on the metal-enriched volume fraction by the outflows. 
In a forthcoming paper, we will focus on the metal abundances in the IGM
resulting from anisotropic AGN outflows, and their observational
consequences.

\acknowledgments 

We thank Philip Hopkins, Matthew Pieri, and Paul Wiita 
for useful correspondence. 
The subroutine that calculates 
the direction of least resistance was written by C\'edric Grenon.
All calculations were performed at the Laboratoire 
d'astrophysique num\'erique, Universit\'e Laval. 
We thank the Canada Research Chair program and NSERC for support.

\appendix

\section{NUMERICAL SOLUTION FOR THE OUTFLOW}

In this appendix, we describe the technique used for solving the
equations describing the evolution of the outflows. The presentation
closely follows the one used in the appendix of PMG07.

The equations governing the evolution of the outflow are
\begin{eqnarray}
\label{ddotR}
\ddot R  &=& {4\pi R^2[1-\cos(\alpha/2)]\over M_s}
             (p^{\phantom1}_T+p^{\phantom1}_B-p^{\phantom1}_x) 
             -{G\over R^2}\left[M_{\rm gal}+{M_s\over2}\right]
             -{\Omega(z)H(z)^2R\over2}
              \nonumber \\
         && +\Omega_\Lambda(z)H^2(z)R -{\dot M_s\over M_s}[\dot R-v_p(R)]\,, \\
\label{dotMs}
\dot M_s &=& \cases{0\,, & $v_p(R)\geq\dot R$, \cr\displaystyle
                 {3\Omega_b(z)H^2(z)R^2\over2G}
                 \left(1-\cos{\alpha\over2}\right)[\dot R-v_p(r)]\,, 
                         & $v_p(R)<\dot R$, \cr} \\
\label{dotpT}
\dot p^{\phantom1}_T &=& {\Lambda\over2\pi R^3[1-\cos(\alpha/2)]}
-5p^{\phantom1}_T{\dot R\over R}\,,\\
\label{dotpB}
\dot p^{\phantom1}_B &=& {\epsilon^{\phantom1}_BL_{\rm AGN}
\over4\pi R^3[1-\cos(\alpha/2)]}
-4p^{\phantom1}_B{\dot R\over R}\,.
\end{eqnarray}

\noindent with the initial conditions $R=0$ at $t=0$. 
We used equations~(\ref{rhox}) and (\ref{Md}) to eliminate $M_d$ and $\rho_x$
in equations~(\ref{ddotR}) and (\ref{dotMs}).
Notice that
since $M_s\propto1-\cos(\alpha/2)$, the dependencies of the opening
angle $\alpha$ cancel out in the first and last terms of
equation~(\ref{ddotR}). The only remaining term in that equation
that depends on $\alpha$ is the term $-GM_s/2R^2$. The most important
dependencies on $\alpha$ are found in equations~(\ref{dotpT}) and
(\ref{dotpB}). The energies (thermal and magnetic) 
are injected into smaller volumes, leading to larger pressures.

Many terms in 
equations~(\ref{ddotR})--(\ref{dotpB}) diverge at $t=0$, making it impossible
to find a numerical solution in this form (these divergences occur because
we neglect the finite size of the source; in the real universe, many terms
are large, but finite, at a radius $R$ that it small, but nonzero). To improve
the behavior of the equations at early time, we perform a change of variable.
To find the appropriate change of variable, we take the limit $t\rightarrow0$
in equations~(\ref{ddotR})--(\ref{dotpB}), and keep only the leading terms.
We get
\begin{eqnarray}
\label{ddotR0}
\ddot R  &=& {4\pi R^2[1-\cos(\alpha/2)]
(p^{\phantom1}_T+p^{\phantom1}_B)\over M_s}
             -{\dot M_s\dot R\over M_s}\,,\\
\label{dotMs0}
\dot M_s &=& 4\pi R^2[1-\cos(\alpha/2)]\rho_{x,i}\dot R\,, \\
\label{dotpT0}
\dot p^{\phantom1}_T &=& {\Lambda_i\over2\pi R^3[1-\cos(\alpha/2)]}
-5p^{\phantom1}_T{\dot R\over R}\,,\\
\label{dotpB0}
\dot p^{\phantom1}_B &=& {\epsilon^{\phantom1}_B\Lambda_i
\over4\pi R^3[1-\cos(\alpha/2)]}
-4p^{\phantom1}_B{\dot R\over R}\,.
\end{eqnarray}

\noindent 
where the subscripts $i$ indicates initial values at $t=0$.
Notice that in equation~(\ref{dotpB0}), we used $L_{\rm AGN}=\Lambda$ in the
limit $t\rightarrow0$.
We can easily show that the solutions of 
equations~(\ref{ddotR0})--(\ref{dotpB0}) are power laws,
\begin{eqnarray}
\label{R0}
R  &=& Ct^{3/5}\,,\\
\label{Ms0}
M_s &=& {4\pi C^3[1-\cos(\alpha/2)]\rho_{x,i}\over3}t^{9/5}\,, \\
\label{pT0}
p^{\phantom1}_T &=& {7C^2\rho_{x,i}
\over25(1+\epsilon^{\phantom1}_B/2)}t^{-4/5}\,,\\
\label{pB0}
p^{\phantom1}_B &=& {7C^2\rho_{x,i}
\over25(1+2/\epsilon^{\phantom1}_B)}t^{-4/5}\,,
\end{eqnarray}

\noindent where
\begin{equation}
\label{eq-C}
C=\left\lbrace{125\Lambda_i(1+\epsilon^{\phantom1}_B/2)
\over154\pi\rho_{x,i}[1-\cos(\alpha/2)]}\right\rbrace^{1/5}\,.
\end{equation}

Using equations~(\ref{R0})--({\ref{pB0}), we can find
the proper change of variables. We define
\begin{eqnarray}
\label{linS}
S   &=&Rt^{2/5}\,,\\
\label{linNs}
N_s &=&M_st^{-4/5}\,,\\
\label{linqT}
q^{\phantom1}_T &=&p^{\phantom1}_Tt^{9/5}\,,\\
\label{linqB}
q^{\phantom1}_B &=&p^{\phantom1}_Bt^{9/5}\,,\\
\label{linU}
U &=&\dot St\,,\\
\label{linO}
O_s &=& \dot N_st\,.
\end{eqnarray}

\noindent In the limit $t\rightarrow0$, the six functions $S$, $N_s$,
$q^{\phantom1}_T$, $q^{\phantom1}_B$, $U$, and $O_s$ vary linearly with
$t$. We now eliminate the functions $R$, $M_s$, $p^{\phantom1}_T$ and
$p^{\phantom1}_B$ in our original equations~(\ref{ddotR})--(\ref{dotpB})
using equations~(\ref{linS})--(\ref{linO}) and get
\begin{eqnarray}
\label{dotU}
\dot U  &=& {9U\over5t}-{14S\over25t}+{4\pi S^2
             (q^{\phantom1}_T+q^{\phantom1}_B-q^{\phantom1}_x)\over N_st^2}
             \left(1-\cos{\alpha\over2}\right)
             -{GM_{\rm gal}t^{11/5}\over S^2}-{\Omega H^2St\over2}
              \nonumber \\ &&
             -{GN_st^3\over2S^2}
        +\Omega_\Lambda H^2St
        -\left[{O_s\over N_s}+{4\over5}\right]
         \left[{U\over t}-{2S\over5t}-HS\right]\,, \\
\label{dotNs}
\dot N_s &=& \cases{\displaystyle-{4N_s\over5t}\,, & 
                    $\displaystyle HSt+{2S\over5}\geq U$, \cr
\displaystyle{3\Omega_bH^2S^2\over2G}\left(1-\cos{\alpha\over2}\right)
\left[{U\over t^3}-{2S\over5t^3}-{HS\over t^2}\right]-{4N_s\over5t}\,, 
                    & $\displaystyle HSt+{2S\over5}<U$, \cr} \\
\label{dotqT}
\dot q^{\phantom1}_T &=& {\Lambda t^3\over2\pi S^3[1-\cos(\alpha/2)]}
-{5q^{\phantom1}_TU\over St}+{19q^{\phantom1}_T\over5t}\,,\\
\label{dotqB}
\dot q^{\phantom1}_B &=& {\epsilon^{\phantom1}_BL_{\rm AGN}\,t^3\over
4\pi S^3[1-\cos(\alpha/2)]}
-{4q^{\phantom1}_BU\over St}+{17q^{\phantom1}_B\over5t}\,,\\
\label{dotS}
\dot S &=& {U\over t}\,,
\end{eqnarray}

\noindent where
$q^{\phantom1}_x\equiv p^{\phantom1}_xt^{9/5}$ in equation~(\ref{dotU}).

These equations are completely equivalent to our original 
equations~(\ref{ddotR})--(\ref{dotpB}), but all divergences at $t=0$ have
been eliminated. These equations can therefore be integrated numerically
using a standard Runge-Kutta algorithm, with the initial
conditions $U=S=N_s=q^{\phantom1}_T=q^{\phantom1}_B=0$ at $t=0$.
However, before doing so, it is preferable to rewrite the equations
in dimensionless form. We define
\begin{eqnarray}
\label{dimtau}
\tau                   &\equiv& H_0t\,, \\
\label{dimS}
\tilde S               &\equiv& {H_0S\over C}\,, \\
\tilde U               &\equiv& {H_0U\over C}\,, \\
\tilde N_s             &\equiv& {GN_s\over C^3H_0}\,, \\
\tilde O_s             &\equiv& {GO_s\over C^3H_0}\,, \\
\label{dimqT}
\tilde q^{\phantom1}_T &\equiv& {Gq^{\phantom1}_T\over C^2H_0}\,, \\
\tilde q^{\phantom1}_B &\equiv& {Gq^{\phantom1}_B\over C^2H_0}\,, \\
\tilde q^{\phantom1}_x &\equiv& {Gq^{\phantom1}_x\over C^2H_0}\,, \\
f_H                    &\equiv& {H\over H_0}\,, \\
f_\Lambda              &\equiv& {\Lambda\over\Lambda_i}\,, \\
{\cal M}               &\equiv& {GM_{\rm gal}\over C^3H_0^{1/5}}\,.
\end{eqnarray}

\noindent Equations~(\ref{dotU})--(\ref{dotS}) reduce to their 
dimensionless form,
\begin{eqnarray}
\label{dotUt}
{d\tilde U\over d\tau}  
&=& {9\tilde U\over5\tau}-{14\tilde S\over25\tau}
+{4\pi\tilde S^2
(\tilde q^{\phantom1}_T+\tilde q^{\phantom1}_B-\tilde q^{\phantom1}_x) 
\over\tilde N_s\tau^2}\left(1-\cos{\alpha\over2}\right)
-{{\cal M}\tau^{11/5}\over\tilde S^2}-{\Omega f_H^2\tilde S\tau\over2}
\nonumber \\ &&
-{\tilde N_s\tau^3\over2\tilde S^2}+\Omega_\Lambda f_H^2\tilde S\tau
-\left[{\tilde O_s\over\tilde N_s}+{4\over5}\right]
\left[{\tilde U\over\tau}-{2\tilde S\over5\tau}-f_H\tilde S\right]\,, \\
\label{dotNst}
{d\tilde N_s\over d\tau} 
&=& \cases{\displaystyle-{4\tilde N_s\over5\tau}\,, & 
$\displaystyle f_H\tilde S\tau+{2\tilde S\over5}\geq\tilde U$, \cr
\displaystyle{3f_H^2\Omega_b\tilde S^2\over2}\left(1-\cos{\alpha\over2}\right)
\left[{\tilde U\over\tau^3}-{2\tilde S\over5\tau^3}
-{f_H\tilde S\over\tau^2}\right]-{4\tilde N_s\over5\tau}\,, 
& $\displaystyle f_H\tilde S\tau+{2\tilde S\over5}<\tilde U$, \cr} \\
\label{dotqTt}
{d\tilde q^{\phantom1}_T\over d\tau}
&=& {231f_{Hi}^2\Omega_{bi}f^{\phantom1}_\Lambda\tau^3
\over1000\pi\tilde S^3(1+\epsilon^{\phantom1}_B/2)}
-{5\tilde q^{\phantom1}_T\tilde U\over\tilde S\tau}
+{19\tilde q^{\phantom1}_T\over5\tau}\,,\\
\label{dotqBt}
{d\tilde q^{\phantom1}_B\over d\tau}
&=& {231f_{Hi}^2\Omega_{bi}f^{\phantom1}_\Lambda\tau^3\over
1000\pi\tilde S^3(1+2/\epsilon^{\phantom1}_B)}
-{4\tilde q^{\phantom1}_B\tilde U\over\tilde S\tau}
+{17\tilde q^{\phantom1}_B\over5\tau}\,,\\
\label{dotSt}
{d\tilde S\over d\tau} &=& {\tilde U\over\tau}\,,
\end{eqnarray}

\noindent where $\tilde O_s\equiv\tau(d\tilde N_s/d\tau)$ in 
equation~(\ref{dotUt}). In the limit $\tau\rightarrow0$, where many terms 
take the form $0/0$, the derivatives reduce
to $d\tilde S/d\tau = d\tilde U/d\tau = 1$,
$d\tilde q^{\phantom1}_T/d\tau=21\Omega_{b,i}f^2_{H,i}/200\pi(1+\epsilon_B/2)$,
$d\tilde q^{\phantom1}_B/d\tau=21\Omega_{b,i}f^2_{H,i}/200\pi(1+2/\epsilon_B)$,
and
$d\tilde N_s/d\tau = \Omega_{b,i}f^2_{H,i}\left(1-\cos\alpha/2\right)/2$.

The quantity $f_\Lambda$ appearing in equations~(\ref{dotqTt}) 
and~(\ref{dotqBt}) depends on the luminosity $L_{\rm Comp}$,
which is given by equation~(\ref{eq-Lcomp}).  We eliminate $P_T$ and $R$ 
in equation~(\ref{eq-Lcomp}),
using equations~(\ref{linS}), (\ref{linqT}), (\ref{dimtau}), (\ref{dimS}), 
and (\ref{dimqT}), then eliminate $C$ using
equation~(\ref{eq-C}). We get

\begin{equation}
\label{eq-LcompLi}
{L_{\rm Comp}\over L_i}={200\pi^3\over 2079}
{\sigma_T\hbar\over m_eH_0\Omega_{b,i} f^2_{H,i}}
{\left(kT_{\gamma 0} \over \hbar c\right)^4}{\left(1+z\right)^4}
{\tilde q_T\tilde S^3 \over \tau^3}\,.
\end{equation}

%

\end{document}